\documentstyle[epsfig,amsfonts,amssymb,diagram,11pt]{article}

\makeatletter
\@addtoreset{equation}{section}

\newcommand{\R}{{\Bbb R}}
\newcommand{\C}{{\Bbb C}}

\def\SL{{\rm SL}}
\def\H{{\bf H}}
\def\U{{\bf U}}

\newcommand{\be}{\begin{eqnarray}} 
\newcommand{\ee}{\end{eqnarray}}

\begin{document}

\begin{titlepage}

\thispagestyle{empty}

\title{$\Lambda<0$ Quantum Gravity in 2+1 Dimensions II: \\
Black Hole Creation by Point Particles}

\author{
{\bf Kirill Krasnov}\thanks{{\tt 
krasnov@cosmic.physics.ucsb.edu}}
\\
{\it Department of Physics}\\
{\it University of California, Santa Barbara, CA 93106}}

\date{\normalsize April, 2002}
\maketitle

\begin{abstract}
\normalsize Using the recently proposed formalism for $\Lambda<0$ quantum 
gravity in 2+1 dimensions we study the process of black hole production 
in a collision of two point particles. The creation probability for
a BH with a simplest topology inside the horizon is given by the Liouville 
theory 4-point function projected on an intermediate state. We analyze in 
detail the semi-classical limit of small AdS curvatures, in which the 
probability is dominated by the exponential of the classical Liouville 
action. The probability is found to be exponentially small. We then argue that
the total probability of creating a horizon given by the sum of
probabilities of all possible internal topologies is of order unity,
so that there is no exponential suppression of the total production rate.
\end{abstract}

\end{titlepage}


\section{Introduction}
\label{sec:intr}

In \cite{String} we have proposed a formalism for
quantum theory of asymptotically AdS 2+1 gravity. The formalism 
is based on the analytic continuation procedure developed in
\cite{Cont}. The theory is defined holographically, using 
a conformal field theory (CFT) on the boundary. The relevant
CFT is expected to be the quantum Liouville field theory (LFT),
or, more precisely, a certain extension thereof that 
incorporates the point particle states, see the companion
paper \cite{String} for more details.

In the present paper we apply this formalism to
study black hole (BH) creation by point particles.
Classically such a process was described in two papers
\cite{Matsch,Matsch-Rot}. The process is essentially a
$\Lambda<0$ version of the Gott time machine \cite{Gott}. 
In this paper we obtain the quantum amplitude for black hole 
production. We verify the expected semi-classical behavior, in that
the amplitude is peaked on the classical Matschull process. 

It is important to realize that there is not one but many
different classical Matschull processes. Namely, the
BH created in a collision of two point particles can have
a non-trivial topology inside the horizon. The process
described in \cite{Matsch} is that corresponding to
the simplest possible internal topology. In the main text we give 
examples of other possible internal topologies that can get created. 
The internal structure inside the horizon is invisible to an
external observer, so is correct to think of it as
the BH microstate. The formalism of \cite{String} allows
to calculate the creation probability for each internal
topology (each microstate). The total probability of
creating a horizon is then given by a sum of probabilities
of all different possible microstates. In this paper we only
study in detail the amplitude (probability) for the simplest
microstate. We then give some arguments as to what the
result of the sum over microstates can be.

Thus, the paper mostly deals with the process
described in \cite{Matsch}, which is the simplest
one in which a horizon is created. The calculation of the amplitude 
for this process proceeds as follows. First
we perform an analytic continuation of the classical Matschull
process. The result is a particular hyperbolic 3D manifold $M$. The
topology of the asymptotic boundary $\tilde{X}$ of this manifold 
is that of a sphere with 4 conical singularities. In a certain
precise sense $\tilde{X}$ can be viewed  as the {\it double} of a spatial 
slice $X$ of the spacetime, see below. According to the prescription
\cite{String} the Hartle-Hawking state is given by the CFT
partition function on $X$, which, in our
case is a disc with two conical singularities.
The BH creation amplitude $\Psi$ is thus the LFT 2-point function on 
a disc:
\be
\Psi[a;\eta,\eta]=Z_{\rm LFT}\left[\,\,
\lower0.23in\hbox{\epsfig{figure=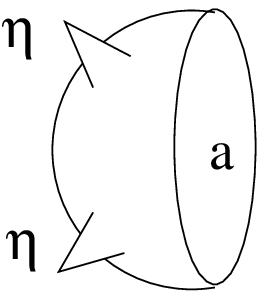,height=0.5in}}
\,\,\right].
\ee
The amplitude depends on the boundary condition $a$, on the
rest mass of the particles $\eta$, and on the relative position
of the insertion points, which encodes the center of mass 
momentum of particles. This 2-point function is explicitly known, see 
\cite{Teschner-B2}. In practice, however,
we are more interested in a probability given by the amplitude squared.
To obtain the probability let us take $|\Psi|^2$ and sum over all
possible boundary conditions. Summing over boundary conditions is
equivalent to ``erasing'' them, and the result is the 4-point
function on the sphere:
\be\label{4-point}
\sum_a |\Psi[a;\eta,\eta]|^2 = Z_{\rm LFT}\left[\,\,
\lower0.3in\hbox{\epsfig{figure=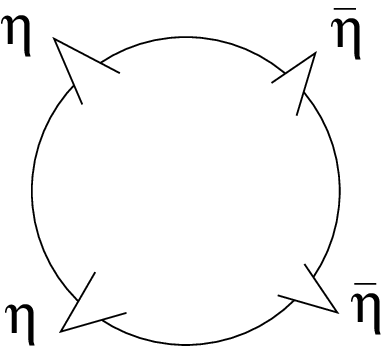,height=0.6in}}
\,\,\right].
\ee
This quantity has the interpretation of the probability of two
point particles colliding and forming a BH. To get 
the probability of creation of a {\it particular size} BH we
have to project the 4-point function on some intermediate
state. We have, schematically,
\be\label{prob}
Z_{\rm LFT}\left[\,\,
\lower0.3in\hbox{\epsfig{figure=4-point.eps,height=0.6in}}
\,\,\right] = \sum_P \,\,
\lower0.3in\hbox{\epsfig{figure=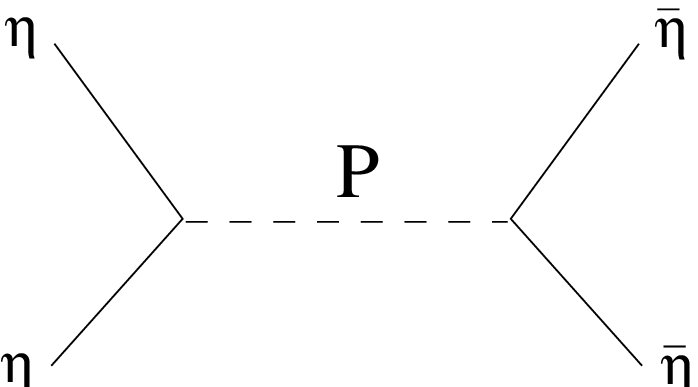,height=0.6in}}
\ee
The sum here is actually an integral, see below. Each term in the sum 
(\ref{prob}) has the interpretation of the probability of
creation of a BH of a particular size determined by the label $P$,
see formula (\ref{size-P}) below for a precise identification. 
Note that the same quantity can also be interpreted
as the scattering amplitude for two point particles with the
intermediate state being a BH. A determination of probability
for production out of the 4-point function is standard in string
theory, see, for example, a recent calculation \cite{Roberto}
of the production rate for string balls. This once more
emphasizes the ``stringy'' nature of this quantum 
theory.

The BH creation probability defined by (\ref{prob}) depends on 
the parameters $\eta$ (rest masses) of
point particles, on the size of the BH being created, and, importantly,
on the cross-ratio of the 4 insertion points on the sphere. 
In \cite{String} we have stated that the cross-ratio $x$
carries information about the relative momentum of the
two colliding particles. When the $x$ is complex its imaginary
part encodes information about the impact parameter. In this paper
we will obtain these relations. As we shall see, this is not easy, 
for these relations, if known, give a solution to the very hard problem
of accessory parameters for uniformization of Riemann surfaces
(in our case a 4-punctured sphere). We shall see that one can
explicitly solve this problem for a particular Riemann surface that 
describes the collision of maximally massive particles. Unfortunately,
this is not the most interesting case physically. However,
a collision of, for example, massless particles, can still be
analyzed in an important limit of a large BH created,
without having an explicit solution for uniformization.

As we explain below, in the semi-classical
limit of small AdS curvatures, the 4-point function is dominated
by the exponential of the classical Liouville action. As it turns out,
for a large BH created, the Liouville action is proportional
to the BH size. Thus, we find that the probability of producing a 
BH of size $2\pi r_+$ with the simplest possible topology inside is 
exponentially small, and is given by:
\be\label{intr-prob}
{\cal P}\simeq e^{-2\pi r_+ /8 G\hbar}.
\ee
This result is not surprising, because this is also
the probability of a large BH exploding into a pair of
point particles, a Hawking type process that is suppressed
by the usual Boltzmann factor. We note that the quantity
in the exponential is equal to $-\beta M/2$, where $\beta$
is the inverse temperature and $M$ is the mass of the BH
of size $2\pi r_+$. The probability (\ref{intr-prob}) should not
be interpreted as the total probability to create a horizon,
instead this is the probability to form a BH in a particular
microstate (particular internal topology). A BH can be formed in 
other microstates. To find the total probability of creating
a horizon one should sum over all possible microstates.

An argument similar to the one in \cite{Voloshin}
can be given to estimate what the result of this sum can be.
Let us take for the moment, as in \cite{Voloshin}, the probability 
of each individual microstate to be $e^{-\beta M}$. The total
number of possible BH microstates is $e^{S_{\rm BH}}$,
where $S_{\rm BH}$ is the BH entropy. The point is that,
unlike for the Schwarzschild BH in 3+1 dimensions, for
BTZ BH the quantity $-I = S_{\rm BH}-\beta M$, where $I$ is the 
BH free energy, is positive: $-I>0$. This is because,
for the BTZ BH, $\beta M=S_{\rm BH}/2$, unlike the
Schwarzschild case $\beta M = 2 S_{\rm BH}$. In other
words, the BTZ BH free energy is negative (like for
a regular thermodynamic system), while for Schwarzschild
it is positive. Thus, the argument of \cite{Voloshin}
applied to the case of 2+1 dimensions tells us that 
we should not expect an exponential suppression of the
BH production rate: the density of states wins over
the creation probability of each individual microstate.
As the result (\ref{intr-prob}) tells us, the probability
of each individual microstate is even larger than
that assumed in the above argument. Thus, there is even less reason
to suspect a suppression in the case of 2+1 gravity.
This is essentially the argument why we believe that
there is no suppression. Some additional arguments
are given in the last section. 

The paper is mostly concerned with the original Matschull
process, which describes creation of a BH of the simplest internal
topology. Other possible topologies appear in section
\ref{sec:cont} and their role is emphasized in \ref{sec:discussion}.

The paper is organized as follows. First, in section \ref{sec:class}
we review the classical Matschull process. We carry out an analytic
continuation of the corresponding spacetimes in section \ref{sec:cont}.
Some basics of quantum LFT are reviewed in section \ref{sec:lft}.
In section \ref{sec:prob} we give an expression for the BH creation 
probability, and study it for the case of maximally massive particles in 
section \ref{sec:massive}. The probability for BH production is 
obtained in section \ref{sec:large}. We conclude with a discussion.

\section{The classical black hole creation process}
\label{sec:class}

In this section we review, in the amount needed for our purposes, the
classical BH creation process. It was initially described for
the case of a head-on collision of massless particles in \cite{Matsch},
and generalized to a non-zero impact parameter in \cite{Matsch-Rot}.
The head-on massive particles case was analyzed in paper 
\cite{Ivo-Formation}.

\subsection{A point particle spacetime}

We start by considering a point particle that at $t=0$ moves through
the origin of AdS${}_3$. A point particle in 2+1 dimensions is described
as a line of conical singularities. This line is a geodesic, it is 
the world-line of the particle in the spacetime, and it is also the 
axis of identifications. In the $\SL(2,\R)$ model of AdS${}_3$,
which is briefly reviewed in the Appendix,
geodesics that pass through the origin ${\bf x}={\bf e}$ are described
as one-parameter subgroups of the type $e^{S{\bf n}}$, where
$s\in\R, {\bf n}\in{\mathfrak sl}(2)$. Let us parameterize the Lie
algebra ${\mathfrak sl}(2)$ in the following way:
\be\label{vf-parametrization}
{\bf n} =
{\bf \gamma}_0 + \kappa {\bf \gamma}(\theta), \qquad \kappa\in\R,
\ee
where $\gamma(\theta)=\gamma_1 \sin(\theta)+\gamma_2\cos(\theta)$.
For a timelike geodesic, which corresponds to a massive particle,
$\kappa<1$. The case of a null particle corresponds to $k=1$.
The exponential can be readily computed:
\be\label{geod}
A = e^{s{\bf n}} =
\cos(\pi\alpha) + {\sin(\pi\alpha)\over\sqrt{1-\kappa^2}}
(\gamma_0 + \kappa\gamma(\theta)),
\ee
where we have introduced
\be\label{def-alpha}
\pi\alpha = s\sqrt{1-\kappa^2}.
\ee
The set of points $e^{S{\bf n}}$ is the axis of the identification 
that has to be carried out to get the particle spacetime. This
identification is generated by the following isometry:
\be\label{isom-non-rot}
{\bf x}\to e^{s{\bf n}}{\bf x} e^{-s{\bf n}}.
\ee
It is clear that all points on the axis are left invariant under
this transformation. 

The spacetime obtained as the result of identifications (\ref{isom-non-rot})
describes a single point particle of a particular mass. The mass
can be obtained from the trace  $(1/2){\rm Tr}(A) = \cos(\pi\alpha)$
of the group element generating the identifications. There are in fact
two different notions of mass in AdS${}_3$. One notion, see, e.g.,
\cite{Welling} arises from considering a dispersion relation.
One can define the particle's momentum as the ${\mathfrak sl}(2)$
element that is the traceless part of $e^{s{\bf n}}$. One then gets:
\be\label{disp-relation}
p^2 + \sin^2{\pi\alpha} = 0.
\ee
It is thus natural to identify the quantity $\pi\alpha$ as the
particle's mass:
\be\label{mass-m}
m = \pi\alpha.
\ee
For small $m$ the dispersion relation (\ref{disp-relation})
is approximately that of a particle in a flat spacetime. 
However, $m$ close to $\pi$ again gives the dispersion relation
appropriate for a small mass particle. The maximum possible
mass corresponds to $m=\pi/2$. This gives the largest possible value
of the norm of the momentum vector. We shall refer to such particles as 
maximally massive. The other notion of mass
is the AdS analog of the usual ADM mass. We shall denote it by $M$ and 
choose it so that the AdS spacetime has zero mass. Then $M$
is related to the parameter $\alpha$ via
\be\label{mass-M}
M/\pi = 1-\alpha^2.
\ee
This relation is explained in the beginning of section \ref{sec:prob}.
As it is clear from this formula, the ``zero'' mass particle $m=0$
has $M=\pi$ and is {\it not} the empty AdS spacetime. As we shall
see later, it is actually the ``zero'' mass BH. One gets the
empty AdS for $m=\pi$, the other ``massless'' case. It is important
to keep in mind this double nature of mass in AdS.

\subsection{Head-on collision}

We are now ready to study particle collisions.
We first consider the case of zero impact parameter. We take 
two point particles, not necessarily massless, thus generalizing the 
analysis of \cite{Matsch}. Such a more general collision has been 
analyzed in \cite{Ivo-Formation}. Note that we use a different 
parameterization of group elements.

For simplicity, we assume that particles have 
same mass. We choose the timelike
geodesics --world-lines of the particles-- 
be generated by the following two vector fields (VF): 
\be
{\bf n}_1 = \gamma_0 +\kappa \gamma_1, \qquad
{\bf n}_2 = \gamma_0 -\kappa \gamma_1, \qquad \kappa\leq 1.
\ee
The case of null particles considered by \cite{Matsch} corresponds to
$\kappa=1$. One gets the generators of identifications by 
exponentiating the above VF's:
\be\nonumber
A_1=e^{s{\bf n}_1} = \cos(\pi\alpha) +{\sin(\pi\alpha)\over\sqrt{1-\kappa^2}}
(\gamma_0 +\kappa \gamma_1), \\
A_2=e^{s{\bf n}_2} = \cos(\pi\alpha) +{\sin(\pi\alpha)\over\sqrt{1-\kappa^2}}
(\gamma_0 -\kappa \gamma_1).
\ee
The parameter $\pi\alpha$ is defined in (\ref{def-alpha}).
The world-lines of both particles intersect the $t=0$ plane at the origin, which
gives a collision.

The result of the collision is either a point
particle or a BH. The mass of the resulting object is determined by 
taking the trace of the product of these $\SL(2,\R)$ elements. We get:
\be\label{BH-holonomy}
\cosh{\nu} = - {1\over 2}{\rm Tr} (A_1 A_2) = - \cos^2(\pi\alpha)
+ {1+\kappa^2\over 1-\kappa^2} \sin^2(\pi\alpha).
\ee
When $\nu$ is real, the result of the collision is a black hole, 
and $\nu$ is the half of its horizon size: $\nu=\pi r_+$. Let us 
note that the relation (\ref{BH-holonomy}) between $\nu$
and the momentum $\kappa$ can be rewritten as:
\be\label{nu}
\cosh(\nu/2)=\sin(\pi\alpha)\cosh(\mu/2).
\ee
Here the parameter $\mu$ is introduced as:
\be\label{mu-two}
\cosh(\mu/2) = {1\over \sqrt{1-\kappa^2}}.
\ee
It is now easy to determine which
configurations of particles result in a BH. As is not
hard to see from (\ref{nu}), a BH is created whenever
$|\sin(\pi\alpha)|\cosh(\mu/2)\geq 1$. One can also rewrite 
this as:
\be\label{BH-condition}
|\cos(\pi\alpha)|\leq\kappa.
\ee
One should think of $\kappa$ as a measure of particle's momentum.
Thus, (\ref{BH-condition}) says that a BH is created for a large
enough relative momentum of particles.

The condition (\ref{BH-condition}) is valid for massive particles only.
The massless case, analyzed in \cite{Matsch}, can be obtained by
taking the limit $\kappa\to 1$. In this case the BH formation
condition reduces to $s\geq 1$, which is the one obtained by Matschull.

As one can see from (\ref{BH-condition}), there is a special value
of particle's mass for which a BH is created for any value of
the momentum $\kappa$. This is when the cosine in (\ref{BH-condition})
equals zero. This corresponds to $m=\pi/2$, which is the
case of maximally massive particles. As it is clear from (\ref{mu-two}),
the BH size is then $2\nu=2\mu$.  

\subsection{Non-zero impact parameter}

An analysis for massless particles was given in \cite{Matsch-Rot}. Here
we present a much simpler treatment based solely on manipulations
with generators. To get the BH angular velocity we use a formula 
derived in \cite{Cont}. Unlike in \cite{Matsch-Rot}, we consider 
particles of arbitrary mass, thus generalizing results of this reference.

Let us first find a parameterization of a particle that at $t=0$ is moving
through a point in AdS${}_3$ some distance away from the origin. We
will later take two such particles, thus producing a non-zero
impact parameter collision. Let us consider a particle that at $t=0$ is at 
a point with coordinates $\rho, \varphi$. 
Using the formula (\ref{point}) of the Appendix it is not hard to
find that the matrix representation of this point is given by:
\be
{\bf x} = 
\cosh{\chi}+\sinh{\chi}\gamma(\varphi) = e^{\chi\gamma(\varphi)}.
\ee
Here 
\be
\chi = \log\left({1+\rho\over 1-\rho}\right)
\ee
is the proper distance from the origin. A geodesic passing through
$\bf x$ can be obtained by a shift. Thus, let us introduce a matrix
${\bf h}=e^{\chi\gamma(\varphi)/2}$. Then a geodesic passing through
$\bf x$ is obtained as a one-parameter group ${\bf h} e^{S{\bf n}} {\bf h}$,
where $\bf n$ is defined in (\ref{vf-parametrization}).
All points on this geodesic are fixed by the isometry
\be
{\bf x}\to A_L {\bf x} A_R^{-1}= 
{\bf h}e^{s{\bf n}} {\bf h}^{-1} {\bf x} {\bf h}^{-1}
e^{-s{\bf n}} {\bf h}.
\ee
Let us calculate the left and right group elements generating this
isometry. A straightforward but lengthy calculation gives
\be\nonumber
A^L = {\bf h}e^{s{\bf n}} {\bf h}^{-1} &=&
\cos(\pi\alpha) + {\sin(\pi\alpha)\over\sqrt{1-\kappa^2}} \Big(
\sinh(\chi)\gamma(\pi/2+\varphi) + \kappa\sinh(\chi)\sin(\theta-\varphi)\gamma_0 
\\ &+&\cosh(\chi)\gamma_0 
+ \kappa\left(\cosh^2(\chi/2)\gamma(\theta) -
\sinh^2(\chi/2)\gamma(2\varphi-\theta)\right)\Big).
\ee
The right generator $A^R$ is obtained by replacing $\chi\to -\chi$.
Let us now take $\theta=\pi/2,\varphi=0$. We get:
\be\label{A-1-left}
A_1^L= 
\cos(\pi\alpha) + {\sin(\pi\alpha)\over\sqrt{1-\kappa^2}} 
\Big((\cosh(\chi)+\kappa\sinh(\chi))\gamma_0+
(\kappa\cosh(\chi)+\sinh(\chi))\gamma_1\Big).
\ee
For $\kappa=1$ (null particle) this becomes $1+s\,e^{\chi}(\gamma_0+\gamma_1)$,
which is the same group element as the one considered in \cite{Matsch-Rot}.
To get the right group element one has to replace $\chi\to-\chi$. One
gets:
\be\label{A-1-right}
A_1^R= 
\cos(\pi\alpha) + {\sin(\pi\alpha)\over\sqrt{1-\kappa^2}}
\Big((\cosh(\chi)-\kappa\sinh(\chi))\gamma_0+
(\kappa\cosh(\chi)-\sinh(\chi))\gamma_1\Big).
\ee

Few remarks are necessary about particle's motion. First of all,
when $\kappa=0$ at $t=0$ the particle is at rest at point $\bf x$.
It will then start falling towards the origin along $\varphi=0$,
where it will have certain momentum. Thus, the case of $\kappa=0$ and
non-zero $\chi$ is the same as that of non-zero $\kappa$
and zero $\chi$, except that in the former case the particle
moves through the origin at $t=\pi/2$. To compare the two cases
we note that, for $\chi=0$, the maximal coordinate distance $\rho_{max}$
that the particle can move away from the origin is:
\be
{2\rho_{max}\over 1+\rho_{max}^2}=\kappa.
\ee
The particle first reaches this point
at $t=\pi/2$. The corresponding proper distance from the
origin is ${\rm Arccosh}(1/\sqrt{1-k^2})=\mu/2$, where
$\mu$ is the same quantity that already appeared in 
(\ref{mu-two}). For $\mu, \chi$ both non-zero we have a particle
that at $t=0$ is the proper distance $\chi$ away 
from the origin, and has momentum $\kappa$ in the
direction orthogonal to the line $\varphi=0$. At time
$t=\pi/2$ the particle is the proper distance $\mu/2$
from the origin. When $\chi>\mu/2$ one gets
the same orbits only that the two parameters
interchange. It is thus enough to consider
only the range of parameters $\chi\leq\mu/2$.

Let us now bring the second particle. We take it to be moving in 
the opposite direction, which corresponds to $\theta=3\pi/2$ and passing 
through the point $\rho, \varphi=\pi$. The corresponding left and right 
generators are 
\be\label{A-2}
A_2^L = \cos(\pi\alpha) + {\sin(\pi\alpha)\over\sqrt{1-\kappa^2}} 
\Big((\cosh(\chi)+\kappa\sinh(\chi))\gamma_0-
(\kappa\cosh(\chi)+\sinh(\chi))\gamma_1\Big), \\ \nonumber
A_2^R = \cos(\pi\alpha) + {\sin(\pi\alpha)\over\sqrt{1-\kappa^2}}
\Big((\cosh(\chi)-\kappa\sinh(\chi))\gamma_0-
(\kappa\cosh(\chi)-\sinh(\chi))\gamma_1\Big).
\ee

The result of the collision can be found by finding the left and right group 
elements $A^{L,R}$ generating the corresponding identification. A 
straightforward calculation gives:
\be \nonumber
\cosh\nu={1\over 2}{\rm Tr}(A^L) &=& -{1\over 2}{\rm Tr}(A_1^L A_2^L) = \\
&-&\cos^2(\pi\alpha) + {\sin^2(\pi\alpha)\over 1-\kappa^2}\left(
(1+\kappa^2)\cosh(2\chi) + 2\kappa\sinh(2\chi)\right), 
\ee
\be\nonumber
\cosh\bar{\nu}={1\over 2}{\rm Tr}(A^R) &=& -{1\over 2}{\rm Tr}(A_1^R A_2^R) = \\
&-&\cos^2(\pi\alpha) + {\sin^2(\pi\alpha)\over 1-\kappa^2}\left(
(1+\kappa^2)\cosh(2\chi) - 2\kappa\sinh(2\chi)\right).
\ee
From here, using simple manipulations, we obtain:
\be\label{nu-rot}
\cosh(\nu/2)=\sin(\pi\alpha)\cosh(\mu/2+\chi), \qquad
\cosh(\bar{\nu}/2)=\sin(\pi\alpha)\cosh(\mu/2-\chi),
\ee
where $\mu$ was defined in (\ref{mu-two}). We thus obtained
the same relations as (\ref{nu}), where $\mu/2$ is replaced
by $\mu/2+\chi$ for the left group element and by $\mu/2-\chi$
for the right one. Now the parameters of the corresponding BH can be 
obtained using the formulas derived in \cite{Cont}. We get:
\be\label{rot-parameters}
\pi r_+ = {1\over 2}(\nu+\bar{\nu}), \qquad
\pi r_+ \Omega = \pi r_- = 
{1\over 2}(\nu-\bar{\nu}).
\ee

Let us first consider the case of two particles of the maximal mass.
We then have $\sin(\pi\alpha)=1$ and, 
thus, as for the head-on case, a BH is created for any value of
the momentum $\kappa$. The BH size is $2 \pi r_+=2\mu$ and the 
angular velocity is $\Omega = 2\chi/\mu$. In other words,
the BH size is determined solely by the momentum $\kappa$, and
is given by the same expression as in the head-on case. The
impact parameter $2\chi$ is equal to a half of the inner
horizon size. Let us also note that when $2\chi=\mu$ an
extremal rotating BH is created. Because $\chi\leq\mu/2$
no naked singularity can be created. 

The described case of particles of the maximal mass is somewhat 
pathological because, as is clear from a comparison of the horizon size
$2\pi r_+=2\mu$ and of the proper distance $\mu$ between the particles at
maximum separation, the horizon exists already at maximum
separation. Thus, the spacetime containing two particles
of maximal mass {\it is} a black hole, for the particles
are always behind the horizon. For zero $\chi$, the two
particles fall towards each other, collide and then hit the
BH singularity. For a non-zero
$\chi$ the particles never hit the singularity, oscillating
between the inner and outer horizons. In particular, for the 
extremal case $\chi=\mu/2$, and the particles are always at the 
same distance from each other, just behind the horizon.

Let us now consider the other limiting case, that of massless particles. 
Then $\kappa\to 1$ and we get:
\be
\cosh(\nu/2) = s\, e^{\chi}, \qquad
\cosh(\bar{\nu}/2) = s\, e^{-\chi}.
\ee
As is clear from this expressions a BH is created whenever 
$s\geq e^{\chi}$. Both the BH size and angular momentum can now be
obtained using (\ref{rot-parameters}). It is instructive to plot
$\pi r_+, \pi r_-$ as functions of $\chi$ at fixed relative momentum
$s$. For large $s$ the BH size is $2\pi r_+=2\log(4s^2)+O(1/s^2)$,
and this is practically unchanged for the whole range of
parameter $\chi\leq \log(s)$, except for a small region
near $\chi=\log(s)$. At extremality the black hole size
drops to $2\pi r_+=2\log(2s^2)$, that is by $2\log{2}$.
The dependence of the $2\pi r_-$ on $\chi$ is practically
linear for the whole range of $\chi$. One gets, sufficiently
far from extremality: $2\pi r_- = 4\chi$. Thus, for a large
relative momentum $s$, and sufficiently far from extremality,
the dependence of the inner, outer horizon radii on the 
parameters of the colliding particles is essentially the
same as in the maximally massive case. Null case gives
something new only for either small momentum $s$, or
close to the extremality. As we shall see, this carries
over into the quantum theory. The situation of large $s$
null particles is described in the quantum theory by the
same formulas as the maximally massive case. 

\section{Analytic continuation}
\label{sec:cont}

Here we carry out an analytic continuation of the
spacetimes of the previous section. We use the procedure
described in \cite{Cont}, although only black hole spacetimes
were treated in this reference. A generalization to point
particles is somewhat non-trivial. The content of this section
is new.

\subsection{Zero impact parameter}

We start by considering a single point particle moving at $t=0$
through the origin of AdS. Later we take two such particles moving
in opposite directions to obtain a zero impact parameter collision. 

In order to analytically continue the particle spacetime we, following
\cite{Rot,Cont}, consider the action of the isometry (\ref{isom-non-rot}) 
on the conformal boundary cylinder $\cal I$. The main idea is to
find and analytically continue the fixed points of this action. 
To this end
we need to find the restriction onto $\cal I$ of the VF $\xi$ generating
(\ref{isom-non-rot}). To get $\xi$ we simply have to
replace the $\gamma$-matrices in (\ref{vf-parametrization}) 
by the VF's $J_i$ (\ref{J-gamma}),
and then use (\ref{vf-boundary}) to restrict the result to the
boundary. One has to do this separately for the right and left VF's.
We get, for the left part:
\be
s{\bf n} = - 2s\left(1 + \kappa(\sin{u}\sin{\theta}-\cos{u}\cos{\theta})
\right)\partial_u = -2s(1-\kappa\cos(u+\theta))\partial_u .
\ee
Let us note that this can be rewritten as:
\be\label{vf-form}
s{\bf n} = - 4s\kappa\sin{\left({u+\theta+i\delta\over 2}\right)}
\sin{\left({u+\theta-i\delta\over 2}\right)} \partial_u=
- 4s\kappa\sin{\left({u-u_p\over 2}\right)}
\sin{\left({u-u_{p'}\over 2}\right)} \partial_u.
\ee
Here we have introduced
\be
1/\kappa = \cos{i\delta}.
\ee
and $u_p=i \delta-\theta, u_{p'}=-i \delta-\theta$.
The full VF becomes:
\be\label{vf-non-rot}
\xi=- 4s\kappa\sin{\left({u-u_p\over 2}\right)}
\sin{\left({u-u_{p'}\over 2}\right)} \partial_u+
4s\kappa\sin{\left({v-v_p\over 2}\right)}
\sin{\left({v-v_{p'}\over 2}\right)} \partial_v,
\ee
where $v_p=-i \delta+2\pi-\theta, v_{p'}= i \delta-\theta$. The factor 
of $2\pi$ in $v_p$ is introduced so that the two terms on the right
hand side have the opposite signs. There is some amount of ambiguity here,
for one can always shift any of the null coordinates by $2\pi$. 
As we shall see, our choice gives the correct
pattern of fixed lines for the case of a null particle.

\begin{figure}
\centerline{\hbox{\epsfig{figure=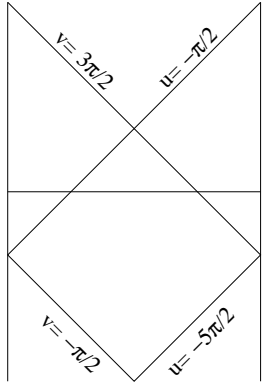,height=2in}}}
\caption{The conformal boundary $\cal I$ for a null particle.
Fixed lines of the VF $\xi$ are shown.}
\label{fig:null-case}
\end{figure}

The two fixed points
of the VF $\xi$ are located at $u_p, v_{p'}$ and $u_{p'},v_p$. In other
words, the $t,\phi$ coordinates are given by:
\be\label{fixed-pts-non-rot}
t = i\delta -\theta, \phi=0,\qquad t'=-i\delta+\pi-\theta, \phi'=\pi.
\ee
Let us see that this gives an expected pattern of fixed lines in the
null case. In this case $\kappa=1$ and $\delta=0$. Let us also take
$\theta=\pi/2$ for definiteness, which corresponds to the particle
1 of the previous section. We have two
fixed points at $t=-\pi/2,\phi=0$ and $t'=\pi/2,\phi'=\pi$.
The corresponding set of fixed lines is shown in Fig.~\ref{fig:null-case}
and is as expected, see \cite{Matsch-Rot}.

\begin{figure}
\centerline{\hbox{\epsfig{figure=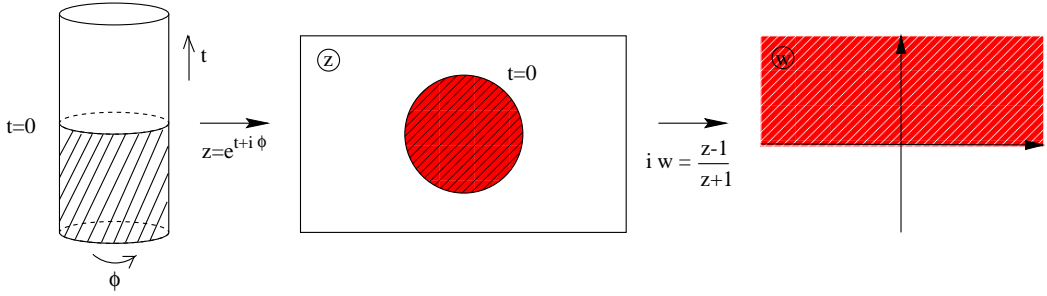,height=1.5in}}}
\caption{Analytic continuation of the boundary cylinder $\cal I$.}
\label{fig:maps-part}
\end{figure}

In case of a massive particle $\kappa<1$ and $\delta\not= 0$. The
coordinates of the fixed points are now complex. This corresponds to
the fact that massive particles never reach the boundary. Indeed,
the fixed points of $\xi$ are exactly those points where 
particle's world-line would intersect the boundary. Let us now
go to the Euclidean cylinder, see Fig.~\ref{fig:maps-part}.
We replace $t\to it$ and map the resulting Euclidean cylinder
to the $z$-plane. Fixed points of $\xi$ go to the following two
points:
\be 
z = e^{-\delta-i\theta}, \qquad z'=e^{\delta+2\pi i-i\theta}.
\ee
From now on we will put $\theta=\pi/2$ for definiteness. We have
$z = -ie^{-\delta}, z'=-ie^\delta$. The corresponding points on the
$w$-plane are:
\be\label{w-1}
w = - {1\over \cosh{\delta}} + i {\sinh{\delta}\over\cosh{\delta}} = 
- \kappa + i \sqrt{1-\kappa^2}={-\sinh(\mu/2)+i\over\cosh(\mu/2)},
\ee
and $w'=1/w=\bar{w}$. We note that both points lie on the unit circle, $w$
above and $w'$ below the real axis. Note that, in case the particle
is not moving $\kappa=0$ and the fixed points are $w=i, w'=-i$. This
is as expected, for these are the points that on the cylinder are
at plus and minus infinity, so that the line connecting them
is the axis of the cylinder. This describes a particle that is located 
at the center of AdS at any $t$. The other limiting case, that
of a null particle, is described by both fixed points approaching
the real axis. 

\begin{figure}
\centerline{\hbox{\epsfig{figure=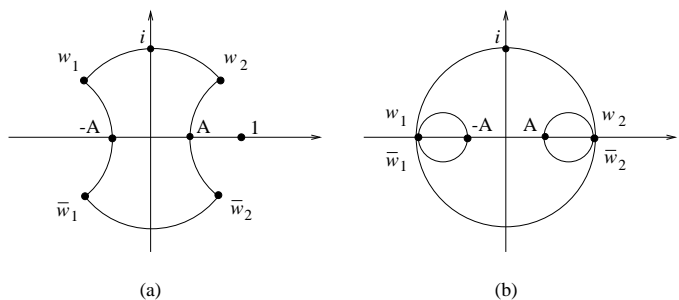,height=1.5in}}}
\caption{A half of the fundamental region for $\Sigma$ is shown:
(a) for the case of particles of the maximal mass $m=\pi/2$;
(b) for the null case.}
\label{fig:fund-region}
\end{figure}

One can repeat the above continuation procedure for the
particle 2 of the previous section, for which $\theta=3\pi/2$.
As one can convince oneself, to go from particle 1 to
particle 2 it is enough to replace $i\delta\to i\delta+\pi$
in formulas for the fixed points. Let us 
denote the fixed points of the first particle by $w_1, w_1'$,
with $w_1$ given by (\ref{w-1}) and those of the second particle
by $w_2, w_2'$. We take $w_2$ to be in the upper half-plane.
We get:
\be\label{w-2}
w_2 = \kappa+i\sqrt{1-\kappa^2}={\sinh(\mu/2)+i\over\cosh(\mu/2)},
\ee
and $w_2'=1/w_2=\bar{w}_2$. All 4 fixed points are located on the unit circle,
see Fig.~\ref{fig:fund-region}.

The analytic continuation of the spacetime in question
can now be described as follows. It is a particular hyperbolic
3D manifold $M$, whose conformal boundary has the topology of
a sphere with 4 conical singularities. $M$ is obtained as
the quotient of the hyperbolic space $\H$ by a certain discrete group
of isometries $\Sigma$. For the case in hand, $\Sigma$ is
a group freely generated by two elements $L_1, L_2\in\SL(2,\C)$.
According to the analytic continuation rule \cite{Cont},
the elements $L_1, L_2$ must be chosen in such a way that
their fixed points coincide with the obtained analytically
continued fixed points $w_1, w_1', w_2, w_2'$. 
The trace does not change in the analytic continuation.
For the considered case of particles of equal mass, and thus same
parameter $\alpha$, see (\ref{def-alpha}), $L_1, L_2$ belong to the
same conjugacy class:
\be\label{trace-L}
{1\over 2}{\rm Tr}(L_1) = {1\over 2}{\rm Tr}(L_2)=\cos(\pi\alpha).
\ee
Since the fixed points of $L_1$ $w_1, w_1'$ are the complex conjugates
of each other (same for $L_2$), the group $\Sigma$ is a discrete subgroup of
$\SL(2,\R)\in\SL(2,\C)$. 

To understand the geometry of $M$ let us describe 
the conformal boundary $\tilde{X}$ of $M$. We recall that the conformal
boundary of the hyperbolic space $\H$ is the Riemann sphere, where
the isometry group $\SL(2,\C)$ acts by fractional linear transformations.
The geometry of the boundary of $\H/\Sigma$ is thus that of ${\cal C}/\Sigma$
where $\cal C$ is the complement of the set of fixed points of
the action of $\Sigma$ on the complex plane. To visualize this
geometry it is enough to find the fundamental region. 
As is not hard to convince oneself, a half of the fundamental
region for the group generated by $L_1, L_2$ is given by
a domain $D$ bounded by 4 circular arcs. The vertices of the
curvilinear polygon $D$ are at the fixed points 
$w_1, \bar{w}_1, w_2, \bar{w}_2$. The angles at all the vertices are
$\pi\alpha$. The boundary $\tilde{X}$ of $M$
can then be pictured as two copies of $D$ glued along their boundaries
to form a sphere with 4 conical singularities. 

The geometry of the boundary is particularly simple when we take
particles of the maximal mass $m=\pi/2$. The angles at all vertices
become right, and $D$ is a polygon bounded by 4 mutually orthogonal 
arcs, see Fig.~\ref{fig:fund-region}(a). As is not hard to find, the 
radius of the two circles bounding $D$ on left and right is
$\sqrt{1-k^2}/k$, the location of the centers on the real axis is at $\pm 1/k$.

For null particles all 4 fixed points lie on the real axis,
at points $\pm 1$. The generators $L_1, L_2$ become parabolic
in this case. A half of the fundamental region is
shown in Fig.~\ref{fig:fund-region}(b).

To further understand the geometry of $\tilde{X}$, let us restrict our 
attention to the upper half-plane $\U$. The group 
$\Sigma\subset\SL(2,\R)$ and thus acts by isometries on $\U$.
A half of the fundamental region for this action is just the
part of $D$ that lies above the real axis. We thus see that
$X=\U/\Sigma$ is a hyperbolic manifold with 2 conical singularities
and one asymptotic region. The circumference $l$ of the throat of 
this asymptotic region can be easily computed. Half of $l$ is given by
the distance in $\U$ between the arcs $w_1, -A$ and $A, w_2$. As
is not hard to find, $l/2=\nu$, where $\nu$ was defined in
(\ref{nu}). This means that the 2D hyperbolic manifold
$X=\U/\Sigma$ has the geometry of 2 point particles inside the
throat of size $l = 2\pi r_+$. In other words, this is the geometry of 
two point particles and a BH of size $2\pi r_+$.

\begin{figure}
\centerline{\hbox{\epsfig{figure=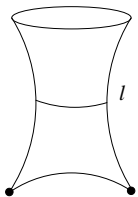,height=1in}}}
\caption{The geometry of $\U/\Sigma$ is that of an infinite throat
ending on two conical singularities. What is shown here is the
geometry for the case of particles of mass $m<\pi/2$. For the
maximal mass the conical singularities are located exactly on
the minimal curve $l$. In the case of null particles one gets
two infinite throats of zero circumference instead of conical
singularities.}
\label{fig:t0-plane}
\end{figure} 

As it became clear from the above discussion, the geometry 
of $\tilde{X}$ is that of
the {\it Schottky double } of $X=\U/\Sigma$. The geometry of $X$
is always that of a BH with two point particles.
The size of the BH can be computed either by multiplying
the holonomies in the Lorentzian signature, as we did in 
the previous section, or by computing the size of the throat
on $\U/\Sigma$. These two different calculations give the
same result. According to the prescription of \cite{String}
the amplitude of a spatial slice geometry $X=\U/\Sigma$ is given
by the LFT partition function on this Riemann surface.
As we discussed in the introduction, the probability
is given by the LFT partition function on the double
$\tilde{X}={\cal C}/\Sigma$, which is a sphere with 4 conical 
singularities. To evaluate this partition function we
will need to understand a relation between the 
geometry of $\tilde{X}$ and location of
the points of insertion of vertex operators in the 4-point function. We will
deal with this below, after we generalize
the analytic continuation to the case of a non-zero
impact parameter.

\subsection{Rotating case}

We now take two particles that are moving towards each other
with some non-zero impact parameter. The analytic continuation
procedure is the same as before. We need to find the VF's that generate
identifications producing the spacetime, find the
restriction of these VF's to the conformal boundary
$\cal I$, and determine the fixed points. The fixed points
are then to be analytically continued by maps in Fig.~\ref{fig:maps-part}
to the complex $w$-plane. Having obtained the fixed points 
$w_1, \bar{w}_1, w_2, \bar{w}_2$ on the $w$-plane one can construct a
group $\Sigma\subset\SL(2,\C)$ generated by, in our case, two 
generators, whose fixed points are $w_1, \bar{w}_1, w_2, \bar{w}_2$. 
The generators are elliptic, corresponding to rotations on an angle
$2\pi\alpha$. This specifies the group $\Sigma$ completely.

In practice, however, it is much easier to simply guess what the locations
of the fixed points must be. Recall, see (\ref{nu}), (\ref{nu-rot}), that
to go from to the non-zero impact parameter case one had to replace
$\mu/2$ by $\mu/2\pm\chi$. Recall also that $\chi$ is related to the
BH angular momentum, and the later should be analytically
continued as we analytically continue the spacetime. This leads
to the following guess for the fixed points:
\be\label{w-1-rot}
w_1 = {-\sinh(\mu/2+i\chi)+i\over\cosh(\mu/2+i\chi)}, \qquad w_1'=1/w_1,
\ee
and
\be\label{w-2-rot}
w_2 = {\sinh(\mu/2+i\chi)+i\over\cosh(\mu/2+i\chi)}, \qquad w_2'=1/w_2.
\ee
This guess for the fixed points can be checked, for example, by
analytically continuing the resulting generators $L_1, L_2$ {\it back}
to the Lorentzian signature, as described in \cite{Cont}. One indeed
gets the isometries $A_1, A_2$ given by (\ref{A-1-left}), (\ref{A-1-right}),
(\ref{A-2}). Note that the fixed points are no longer on the unit
circle, and they are not complex conjugates of each other,
although it is still true that $w'=1/w$.

One can now construct a group $\Sigma$ generated by two elliptic
elements $L_1, L_2$ with traces satisfying (\ref{trace-L}), and
with fixed points given by $w_1, w_1'$ and $w_2, w_2'$ correspondingly.
The group $\Sigma$ is no longer a subgroup of $\SL(2,\R)$.
This is, of course, as expected, for the presence of rotation
is generally manifested by $\Sigma$ being complex.
The space $\H/\Sigma$ is a hyperbolic 3-manifold, whose conformal
boundary $\tilde{X}$ has the topology of a sphere with 4 conical
singularities. The boundary $\tilde{X}$ can be thought of as
obtained by gluing two copies of the non-rotating case
surface $X$ with a twist. We will not need any further details
on the rotating case. As we saw, a non-zero impact parameter
can be incorporated by simply replacing $\mu/2$ by $\mu/2+i\chi$
in all the formulas. Thus, the rotating black hole situation is 
obtained from the zero impact parameter case by an analytic
continuation in $\mu$.

\subsection{Other topologies}

The purpose of this subsection is to note that, in addition to the
simplest topology shown in Fig.~\ref{fig:t0-plane}, more
complicated topologies of the spatial slice may appear
as a result of the collision. The corresponding
spacetimes may be obtained by choosing a more non-trivial
group $\Gamma$.

\begin{figure}
\centerline{\hbox{\epsfig{figure=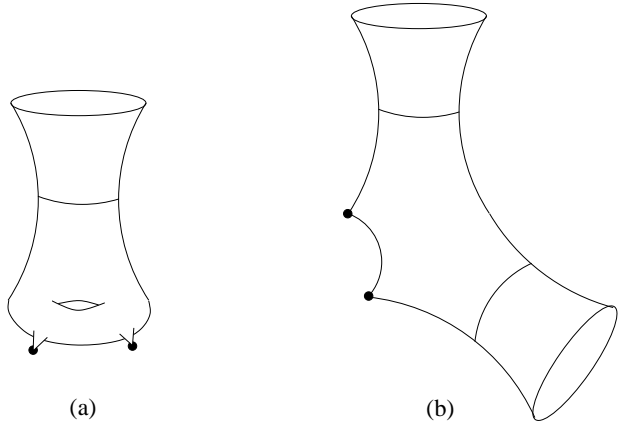,height=2.2in}}}
\caption{Other topologies that can be created inside the
horizon include: (a) a torus; (b) another asymptotic region.}
\label{fig:t0-other}
\end{figure} 

For example, let us consider a group $\Gamma$ generated
by two elliptic elements, corresponding to the two
colliding particles, and, in addition, two hyperbolic
elements whose axes intersect. One gets a spacetime whose
spatial slice topology is depicted in Fig.~\ref{fig:t0-other}(a).
Thus, this spacetime describes a process in which two 
point particles collide and form a BH with a torus inside
the horizon.

Another spacetime is obtained by taking, in addition to the
elliptic particle generators, two hyperbolic generators whose
axes do not intersect. One obtains a spacetime with the
spatial slice topology shown in Fig.~\ref{fig:t0-other}(b)
This is the spacetime describing two colliding particles
that produce a BH with an asymptotic region inside.

It is now clear that one can construct spacetimes describing
two colliding particles producing a BH with an arbitrary
topology inside the horizon. All this topologically
non-trivial structure is behind the horizon, and is thus
invisible from the asymptotic region where particles lived
before the collision. This structure is, in some sense,
a micro-state of the BH created. It is clear that in
order to find the probability of two particles forming 
a horizon one should sum over all possible topologies
that can appear inside. We discuss such a sum in the last 
section.

We are now basically ready to define and study the BH creation
probability. We do this in section \ref{sec:prob}, after
we review some basics of Liouville theory.

\section{Quantum Liouville theory}
\label{sec:lft}

In this section we review some basic facts about the quantum LFT, in
the amount we need for the following. Our main reference here is
\cite{Zamolo}, see also \cite{Teschner-Liouv} and \cite{Takht-Elliptic}.

\subsection{General properties}

The Liouville theory action is:
\begin{equation}
\label{action-L}
A_L[\phi] = 
\int d^2x \left[ {1\over 4\pi} (\partial_a \phi)^2 + \mu e^{2b\phi} \right].
\end{equation}
Here $\phi$ is the Liouville field, $b$ is the coupling constant of the
theory and $\mu$ is a constant of the dimension of $1/({\rm length})^2$,
which sets a scale for the theory. We leave the region over which the integral 
is taken unspecified for now. 
Usually one also adds to the Lagrangian a term proportional to $\phi R$, where 
$R$ is the curvature scalar of a fixed background metric, and 
adjusts the coefficient in front of this term so that the action is
independent of the background. For our purposes, however, it is
more convenient to work in the flat background (this is also the
choice of  \cite{Zamolo}). Then the integral of $\phi R$ translates into 
a set of boundary terms, see below. Although by appropriately choosing the
boundary terms one can define LFT on any Riemann surface, see \cite{Takht}, 
in the present paper we are interested in LFT on the sphere. It is
most convenient to work with the extended complex plane. The field $\phi$ is 
then required to have the following asymptotic for $|z|\to\infty$:
\begin{equation}
\label{b-cond}
\phi(z,\bar{z}) = - Q \log(z\bar{z}) + O(1).
\end{equation}
Here $Q$ is given by:
\begin{equation}
\label{Q}
Q = b + {1\over b}.
\end{equation}
To make the action (\ref{action-L})
well-defined on such fields, one introduces a large disc $D$ of radius
$R\to\infty$ and adds a boundary term to the action:
\begin{equation}
A_L = 
\int_\Omega d^2x \left[ {1\over 4\pi} (\partial_a \phi)^2 + 
\mu e^{2b\phi} \right]+
{Q\over \pi R}\int_{\partial\Omega} dl \, \phi + 2Q^2 \log{R}.
\end{equation}
The last term is needed to make the action finite as $R\to\infty$.

The vertex operators of LFT are:
\begin{equation}
V_\zeta(z) = e^{2\zeta\phi(z)},
\end{equation}
where $z$ is a point on $S^2$. They are primary 
operators of conformal dimension:
\begin{equation}
\Delta_\zeta = \zeta(Q-\zeta).
\end{equation}

Correlation functions of vertex operators are formally defined
as the following functional integral:
\begin{equation}
\label{corr}
{\cal G}_{\zeta_1,\ldots,\zeta_n}(z_1,\ldots,z_n) =
\int {\cal D}\phi \, V_{\zeta_1}(z_1)\cdots V_{\zeta_n}(z_n)
e^{-A_L[\phi]}.
\end{equation}
The integral must be taken over the fields satisfying the boundary 
condition (\ref{b-cond}).

The scale dependence of correlators is:
\begin{equation}
{\cal G}_{\zeta_1,\ldots,\zeta_n}(z_1,\ldots,z_n) =
(\pi\mu)^{(Q-\sum\zeta_i)/b} F_{\zeta_1,\ldots,\zeta_n}(z_1,\ldots,z_n),
\end{equation}
where $F_{\zeta_1,\ldots,\zeta_n}(z_1,\ldots,z_n)$ is independent
of the scale $\mu$. 

The spectrum of LFT consists of the states created by $V_\zeta$ with 
\begin{equation}
\label{spec}
\zeta = {Q\over2} + iP.
\end{equation}
These are the normalizable states. One can also consider the ``states''
created by $V_\zeta$ with $0<\zeta<Q$. These operators create
conical singularities and thus correspond to non-normalizable states.

\subsection{Three-point function}

The three-point function in LFT is given by the 
Dorn-Otto-Zamolodchikov-Zamolodchikov (DOZZ) formula:
\be
{\cal G}_{\zeta_1,\zeta_2,\zeta_3}(z_1,z_2,z_3) = 
|z_{12}|^{2\Delta_{12}} |z_{23}|^{2\Delta_{23}} |z_{31}|^{2\Delta_{31}}
C(\zeta_1,\zeta_2,\zeta_3),
\ee
where $z_{ij}=z_i-z_j, \Delta_{ij}=\Delta_k-\Delta_i-\Delta_j,
i\not=k, j\not=k, \Delta_i=\Delta_{\zeta_i}$ and
\be\nonumber
C(\zeta_1,\zeta_2,\zeta_3) &=& \left[
\pi\mu\gamma(b^2) b^{2-2b^2}\right]^{(Q-\sum\zeta_i)/b} \times 
\\ \label{3-point}
&\times&
{\Upsilon_0 \Upsilon(2\zeta_1)\Upsilon(2\zeta_2)\Upsilon(2\zeta_3)\over
\Upsilon(\zeta_1+\zeta_2+\zeta_3-Q)\Upsilon(\zeta_1+\zeta_2-\zeta_3)
\Upsilon(\zeta_2+\zeta_3-\zeta_1)\Upsilon(\zeta_3+\zeta_1-\zeta_2)}.
\ee
Here 
\be\label{gamma}
\gamma(x)={\Gamma(x)\over\Gamma(1-x)},
\ee
and $\Upsilon(x)$ is a special function defined by the
following integral representation:
\be\label{upsilon}
\log\Upsilon(x) = \int_0^\infty {dt\over t}\left[
\left({Q\over 2}-x\right)^2 e^{-t} - 
{\sinh^2\left({Q\over 2}-x\right){t\over2}\over 
\sinh{bt\over 2}\sinh{t\over 2b}}\right].
\ee
The quantity $\Upsilon_0$ is defined as:
\be
\Upsilon_0=\left({d\Upsilon(x)\over dx}\right)_{x=0}.
\ee

\subsection{4-point function}

The 4-point function can be reduced to a function of the cross-ratio
\be
x={z_{12} z_{34}\over z_{13} z_{24}}.
\ee
It is thus convenient to put $z_4=0, z_3=x, z_2=1, z_1=\infty$.
One introduces:
\be
{\cal G}_{\zeta_1,\ldots,\zeta_4}(\infty,1,x,0) :=
G_{\zeta_1,\ldots,\zeta_4}(x,\bar{x}).
\ee

The 4-point function $G_{\zeta_1,\ldots,\zeta_4}(x,\bar{x})$
can be decomposed into a sum over intermediate states:
\be\label{4-point-interm}
G_{\zeta_1,\zeta_2,\zeta_3,\zeta_4}(x,\bar{x})=
{1\over2} \int_{-\infty}^{\infty}
C(\zeta_1,\zeta_2,{Q\over2}+iP)C(\zeta_3,\zeta_4,{Q\over2}-iP)
|{\cal F}(\Delta_1,\ldots,\Delta_4,\Delta;x)|^2 dP.
\ee
The function ${\cal F}(\Delta_1,\ldots,\Delta_4,\Delta;x)$
is the so-called conformal block \cite{BPZ}, which
sums up all descendants of a given primary 
state. It depends on the central charge of LFT
\be\label{LFT-charge}
c=1+6Q^2,
\ee
and on the conformal dimension
\be
\Delta={Q^2\over 4}+P^2.
\ee

\subsection{Semi-classical limit}

In the semi-classical limit the LFT correlators
are dominated by the classical Liouville action, which
appears as the $b\to 0$ limit of 
(\ref{action-L}). One then introduces a
new ``classical'' Liouville field:
\begin{equation}
\varphi = 2b\phi,
\end{equation}
The ``quantum'' action (\ref{action-L}) is then:
\begin{equation}
A_L[\phi] = {1\over b^2} S_{Liouv}[\varphi],
\end{equation}
where
\begin{equation}
\label{action-class}
S_{Liouv}[\varphi] = {1\over 8\pi} \int d^2x \left[
{1\over2} (\partial_a \varphi)^2 + 8\pi\mu b^2 e^\varphi \right]
\end{equation}
is the classical Liouville action. There are also some boundary terms to 
be added to this action, see below. Varying the classical action with 
respect to $\varphi$ one finds that locally $\varphi$ satisfies the 
classical Liouville equation:
\begin{equation}
\label{Liouv-eq}
\Delta \varphi = 8\pi\mu b^2 e^\varphi.
\end{equation}
Then the metric $ds^2 = e^\varphi |dz|^2$ is a metric of constant
negative curvature $-8\pi\mu b^2$. 

As we have said, correlation
functions ${\cal G}_{\zeta_1,\ldots,\zeta_n}(z_1,\ldots,z_n)$ 
of vertex operators are dominated, in the semi-classical limit,
by $S_{Liouv}$ evaluated on a particular solution
to the Liouville equation. Let us consider the case of
``heavy'' vertex operators, which is relevant for this paper. 
Let us take $\zeta_i = \eta_i/b$ with $\eta_i<1/2$ of order $O(1)$,
and consider the case $\sum\eta_i > 1$ so that there is 
a solution to (\ref{Liouv-eq}) with negative curvature. There is
then a unique solution $\tilde{\varphi}$ of the 
Liouville equation (\ref{Liouv-eq})
with the following boundary conditions:
\begin{eqnarray}\nonumber
\varphi(z,\bar{z}) = - 2\log{|z|^2} + O(1) \qquad {\rm at}\quad |z|\to\infty\\
\label{asympt}
\varphi(z,\bar{z}) = - 2\eta_i \log{|z-z_i|^2} +
O(1) \qquad {\rm at}\quad z\to z_i.
\end{eqnarray}
The correlation functions ${\cal G}_{\zeta_1,\ldots,\zeta_n}(z_1,\ldots,z_n)$
are then dominated by the classical Liouville action 
\begin{equation}\label{domin}
{\cal G}_{\zeta_1,\ldots,\zeta_n}(z_1,\ldots,z_n) \sim
\exp{\left( -{1\over b^2} S_{Liouv}[\tilde{\varphi}]\right)}
\end{equation}
evaluated on the canonical Liouville field $\tilde{\varphi}$.
The classical Liouville action $S_{Liouv}$ is given by:
\begin{equation}
\label{action-A}
S_{Liouv}[\varphi] = {1\over 8\pi} \int_D dx^2 \left[ {1\over 2} 
(\partial_a\varphi)^2 +8\pi\mu b^2 e^\varphi \right] + \phi_\infty + 2\log{R} - 
\sum_i \left ( \eta_i \varphi_i + 2\eta_i^2 \log{\epsilon_i} \right).
\end{equation}
Here $D$ is a disc of radius $R$ with small discs of radii
$\epsilon_i$ cut out around each of the singularities at $z_i$, and
\begin{equation}
\varphi_i = {1\over 2\pi\epsilon_i} \int_{\partial\Omega_i} dl \, \varphi.
\end{equation}

\subsection{Semi-classical limit and Riemann surfaces}

A relation to Riemann surfaces and uniformization arises because
the canonical Liouville field $\tilde{\varphi}$ can be obtained if
the uniformization map is known. Consider the Liouville
stress-energy tensor
\be
T_\varphi=\varphi_{zz}-{1\over 2}\varphi_z^2
\ee
for the canonical field $\tilde{\varphi}$. Then
$T_{\tilde{\varphi}}$ is a meromorphic function on 
$\C$ with second order poles at points $z_i$, see, e.g., \cite{Takht-Elliptic}.
One can then consider the Fuchsian differential equation:
\be\label{Fuchs-eq}
{d^2u\over dz^2}+{1\over 2}T_{\tilde{\varphi}}(z) u = 0.
\ee
It is then a classical result, see references in \cite{Takht-Elliptic},
that the monodromy group $\Gamma$ of this equation is, up to conjugation in
$\SL(2,\C)$, a subgroup of $\SL(2,\R)$. It is a discrete
subgroup if and only if the angle deficits are rational,
in other words if and only if $\eta_i=1/2-1/(2l_i)$, where
$l_i$ are positive integers or $\infty$. There is thus a
particular Riemann surface $\tilde{X}=\U/\Gamma$ associated with
every solution of (\ref{Liouv-eq}), and thus with 
every configuration of points $z_i$ on $\C$ labelled by $\eta_i$.
The topology of this Riemann surface is that of a sphere with
$n$ conical singularities of angle deficits $4\pi\eta_i$.

Let us now consider the ratio $f=u_1/u_2$ of two linearly independent 
solutions of the Fuchs equation (\ref{Fuchs-eq}). It is a multi-valued
meromorphic function on $\C$ with ramification points at $z_i$. The
ratio $f=u_1/u_2$ has the property that its Schwarzian derivative 
${\cal S}(f;z)$ is equal to the quadratic differential 
$T_{\tilde{\varphi}}(z)$. This then implies that one can reconstruct
the canonical Liouville field if one knows $f(z)$:
\be\label{Liouv-field}
e^\varphi = {|f'(z)|^2\over ({\rm Im}\,f)^2}.
\ee
Here the ratio $f=u_1/u_2$ is assumed to be normalized in such
a way that the monodromy group $\Gamma$ is in $\SL(2,\R)$.

To summarize, there is a connection between Riemann surfaces,
their uniformization, and solutions of the Liouville equation
(\ref{Liouv-eq}). In particular, there is a one-to-one correspondence
between solutions $\tilde{\varphi}$ and Riemann surfaces. The
LFT correlators are thus dominated, in the semi-classical
regime, by a particular Riemann surface, whose shape 
can be determined by solving the Liouville equation,
finding the quadratic differential $T_{\tilde{\varphi}}(z)$
and then determining the monodromy group of the
corresponding Fuchsian equation. 

\section{BH creation probability and the semi-classical limit}
\label{sec:prob}

In the introduction we have motivated the following prescription for 
calculating the BH creation probability $\cal P$. It is 
given by the 4-point function (\ref{4-point-interm})
projected onto a particular intermediate state:
\be\label{prob*}
{\cal P}(P,x) = 
C(\zeta_1,\zeta_2,{Q\over2}+iP)C(\zeta_1,\zeta_2,{Q\over2}-iP)
|{\cal F}(\Delta_1,\Delta_2,\Delta_1,\Delta_2,\Delta;x)|^2.
\ee
The parameter $P$ labelling the intermediate primary state should be
thought of as a measure of the size of the BH created. The probability 
$\cal P$ is also a function of the cross-ratio $x$ of the 
4 points where the vertex operators are inserted. We will
study the BH creation probability (\ref{prob*}) in the
semi-classical limit, which, as we shall presently see, 
corresponds to small AdS curvatures.

\subsection{Semi-classical limit relations}

A precise relation between the label $P$ and the BH size can be stated
in the semi-classical limit and is as follows.
Let us now restore the dependence of all quantities on $l, G, \hbar$.
There is then the following relation, see \cite{Brown-Hen},
between the mass $M$ of an object and the conformal dimension:
\be
Ml = \Delta+\bar{\Delta}.
\ee
For primary states $P=p/b$ the conformal dimension is real and we get,
in the semi-classical limit, $Ml=(1/2b^2)(1+4p^2)$. We should now 
recall the Brown-Henneaux value of the central charge:
\be
c_{\rm BH}={3l\over 2G\hbar} = {3l\over 2l_p}
\ee
and match it to the LFT central charge (\ref{LFT-charge})
in the semi-classical limit. This gives:
\be\label{b}
{1\over b^2} = {l\over 4l_p}.
\ee
Thus, small AdS curvatures $l>> l_p$ correspond to small $b$,
which is the semi-classical limit of LFT.
In this limit we get, for the mass:
\be\label{mass-p}
M = {1\over 8l_p}\left(1+4p^2\right).
\ee
This should be compared with the usual BTZ BH relation between
the size and mass:
\be\label{mass-r-plus}
M={1\over 8l_p}\left(1+r_+^2\right)
\ee
which gives:
\be\label{size-P}
r_+=2p.
\ee
Therefore, $\nu=\pi r_+=2\pi p$. Having in mind this identification
between $P=p/b$ and $\nu$ we will sometimes write the semi-classical 
BH creation probability as a function of $\nu$ instead of $P$.

In the system of units that we used so far $8\pi G=\hbar=l=1$. In these
units the equations (\ref{mass-p}), (\ref{mass-r-plus}) become
\be
M/\pi = 1+4p^2. 
\ee
One can repeat a similar analysis for primary states with $\zeta=\eta/b$. 
One gets the relation (\ref{mass-M}) with
\be\label{eta-alpha}
\alpha=1-2\eta.
\ee
This is an obvious relation between the angle $2\pi\alpha$
at the tip of the cone and the angle deficit $4\pi\eta$.

Having found a relation between the size $2\nu$ of the BH created and the
parameter $P=p/b$ of the intermediate primary state, we are ready
to study the probability ${\cal P}(\nu,x)$. As we saw in the previous 
section, in the semi-classical limit the 4-point function 
(\ref{4-point-interm}) is dominated by a particular Riemann surface 
whose topology is that of a sphere with 4 conical
singularities. This Riemann surface has one modulus, which we
choose to be the size of the hole one gets by cutting 
the surface into two 2-punctured discs. We will denote this
size by $2 \nu(x)$ indicating that it depends on the cross-ratio $x$.
It also depends on the conformal dimensions of the vertex operators,
but we suppress this for brevity. The fact that the full 4-point
function is dominated by a particular Riemann surface  
means that the function ${\cal P}(P,x)$ will be peaked 
at a particular value of $P$ for a fixed $x$. 
This value of $P$ is such that $P=p/b$ and $2\pi p=\nu(x)$.
In other words, for a fixed cross-ratio, the BH creation
probability is peaked at a BH of size $2\nu(x)$. Now we would
like: (i) find the function $\nu(x)$; (ii) check
that the corresponding BH is the one predicted by the
classical analysis of section \ref{sec:class}; (iii)
find the probability ${\cal P}(\nu, x(\nu))$ as a
function of a size of the BH created. As we shall see,
(i) is a very hard problem, related to the so-called
problem of accessory parameters for uniformization.
We shall present some relevant results for general $\eta$ 
in the next subsection. Further analysis will be given in
section \ref{sec:massive}, where
we specialize to the case of maximally massive 
$\eta=1/4$ particles.

\subsection{Accessory parameters, Schottky uniformization}

The problem we are facing is to find a relation $\nu(x)$ between 
the modulus $\nu$ of the Riemann surface that dominates
the correlator and the cross-ratio $x$. Let us
explain why this problem is equivalent to the
famous problem of accessory parameters. This will
suggest to us a way to get the relation $\nu(x)$ for
a particular case $\eta=1/4$.

In section \ref{sec:lft} we have seen that the semi-classical
limit relation between LFT correlators and Riemann surfaces
goes through the Fuchs equation (\ref{Fuchs-eq}). As we
discussed, every solution $\tilde{\varphi}$ of the classical Liouville
equation leads to a quadratic differential $T_{\tilde{\varphi}}$.
It is the classical result, see, e.g., \cite{Nehari}, Chap. V,
that $T_{\tilde{\varphi}}$ has the following form:
\be\label{accessory}
T_{\tilde{\varphi}} = {1-\alpha_1^2\over 2z^2}&+&
{1-\alpha_2^2\over 2(z-x)^2}+{1-\alpha_3^2\over 2(z-1)^2}\\
\nonumber
&-&{2+\alpha_4^2-\alpha_1^2-\alpha_2^2-\alpha_3^2\over 2z(z-1)}-
{2 x(1-x){\cal C}(x)\over z(z-x)(z-1)}.
\ee
Here $\alpha_i$ are the monodromy parameters at each of
the conical singularities, which we assumed are located 
at points $0,x,1,\infty$ correspondingly. The quantity
${\cal C}(x)$ is the famous accessory parameter, which
is a function of $x$. It is such that 
he monodromy group of the Fuchsian equation (\ref{Fuchs-eq})
is Fuchsian, and determines a Riemann surface in question.
The ratio $f(z)=u_1/u_2$ of two linearly independent solutions
of (\ref{Fuchs-eq}) gives a conformal mapping from the 
complex plane with 4 marked points into a domain in $\U$.
This map is multi-valued, but the corresponding inverse map
$f^{-1}:\U\to \tilde{X}=\C/\{0,x,1,\infty\}$ is single valued. Map
$f$ is called {\it developing}. If known, it allows
to determine $\nu(x)$. It also allows one to find the
Liouville field $\tilde{\varphi}$, see formula
(\ref{Liouv-field}). One can then evaluate
the Liouville action on $\tilde{\varphi}$. As is
known, see \cite{Takht-Elliptic,Italians}
the accessory parameter ${\cal C}(x)$ can be
obtained as the derivative of the Liouville
action $S_{Liouv}[\tilde{\varphi}]$ with
respect to $x$. Thus, one can view the map 
$f$ as the central object. Once it is known,
all other quantities of interest can be determined.

We will find the relation $\nu(x)$ for $\eta=1/4$, and an analog
of the map $f$, by considering the Schottky uniformization
instead of the Fuchsian one. In Schottky uniformization
a Riemann surface is obtained as a quotient of
the complex plane $\C$ (instead of $\U$) with
respect to a discrete group that is a subgroup
of $\SL(2,\C)$ (instead of $\SL(2,\R)$). In
fact, this is what is relevant for our
purposes, for the Riemann surfaces that we
have obtained as the conformal boundary of $M$
were quotients $\C/\Sigma, \Sigma\subset\SL(2,\C)$.
Not every surface with conical singularities
can be uniformized via Schottky. A necessary requirement is
that there is an even number of marked points, and that they can be
paired with the same angle deficit in a pair. Riemann
surfaces obtained as a double of some surface with
boundary can always be uniformized via Schottky. All
Riemann surfaces we encountered in section \ref{sec:cont}
were of this type.

Having a surface that is uniformizable via Schottky
one can pose a problem similar to that we encountered
considering the Fuchs equation. Namely, find a 
map $w(z)$ mapping the surface $\tilde{X}=\C/\{0,x,1,\infty\}$
into a domain in $\C$, the fundamental 
domain of some group $\Sigma$. Given $x$, such a mapping
exists only for a particular $\Sigma$. Knowing this
map one can get the relation $\nu(x)$.

One can complete the maps $f:\tilde{X}\to\U, w:\tilde{X}\to{\cal C}$
by a map $J:\U\to{\cal C}$ relating the Fuchsian and Schottky
uniformization maps. One gets a commutative diagram:
\begin{equation}\label{diagram}
\begin{diagram}
\node{\U} \arrow[2]{e,t}{J} \arrow{se,b}{f^{-1}}
\node[2]{{\cal C}\subset\C} \arrow{sw,r}{w^{-1}} \\
\node[2]{\tilde{X}}
\end{diagram}
\end{equation}
We note that to evaluate the Liouville action on $\tilde{\varphi}$
it is enough to know the map $J$ relating the two uniformizations.
Indeed, if $J$, or rather its inverse $J^{-1}$ is known, one
gets the canonical Liouville field on the Schottky domain via:
\be\label{Liouv-Schottky}
e^{\tilde{\varphi}(w)} = {|J^{-1}_w|^2\over ({\rm Im}\,J^{-1})^2}.
\ee
One can then evaluate the Liouville action directly on the
Schottky domain. This way of finding the Liouville action,
(and of actually defining it) was used in \cite{Takht} for
the case of higher genus surfaces. 

In some cases the problem of finding the map $w$ (and $J^{-1}$)
may be easier than that of finding the Fuchsian uniformization
map $f$. Thus, as we shall see, it is rather easy to
find $w$ for the special case of $\eta=1/4$, that is, of
conical singularities of angle deficit $\pi$. As we will
find in the next section, the map in question is given by 
elliptic functions. We will also see that it is sometimes easier to
evaluate the Liouville action on the Schottky domain. We will use 
this method to get the Liouville action for $\eta=1/4$, and
to get an asymptotic of $S_{Liouv}$ for the null case $\eta=1/2$.

Before we specialize to the case $\eta=1/4$ let us describe another way 
how the relation $\nu(x)$ can be obtained.

\subsection{Extremum of the 4-point function}

Another way to determine the function $\nu(x)$ is to directly look
for an extremum of ${\cal P}(P,x)$ as a function of $P$ at fixed
$x$. Such an analysis was first performed in \cite{Zamolo}, Section 8,
and we will essentially repeat the results of this reference here.

In the semi-classical limit we consider the
case of ``heavy'' particles with parameters $\zeta_i=\eta_i/b$.
As before, we shall consider a collision of particles of equal
rest mass. We thus take $\eta_1=\eta_2=\eta$. A relation 
between $\eta$ and the parameter $\alpha$ is given by (\ref{eta-alpha}).
We take the intermediate state to be $P=p/b$. In the limit
$b\to 0$ the probability is then given by:
\be
{\cal P}(p,x) = C(\eta/b,\eta/b,(1/2+ip)/b)C(\eta/b,\eta/b,(1/2-ip)/b)
|{\cal F}(\Delta,\Delta,\Delta,\Delta,\Delta';x)|^2,
\ee
where
\be 
\Delta={1\over b^2}\eta(1-\eta), \qquad 
\Delta'={1\over b^2}\left({1\over 4}+p^2\right).
\ee
We will now use the fact that both the structure constant
$C$ and the conformal block $\cal F$ are given semi-classically
by certain exponentials, see \cite{Zamolo}. We have:
\be
C(\eta_1/b,\eta_2/b,\eta_3/b)\sim 
e^{-{1\over b^2} S^{(cl)}(\eta_1,\eta_2,\eta_3)},
\ee
where
\be\label{S-cl}
S^{(cl)}(\eta_1,\eta_2,\eta_3)&=&\left(\sum \eta_i -1\right)\log(\pi\mu b^2)
+ F(\eta_1+\eta_2+\eta_3-1)+F(\eta_1+\eta_2-\eta_3)+ \\ \nonumber
&{}& F(\eta_2+\eta_3-\eta_1)+F(\eta_3+\eta_1-\eta_2)-F(0)-F(2\eta_1)-F(2\eta_2)
-F(2\eta_3).
\ee
Here 
\be
F(\eta)=\int_{1/2}^\eta \log{\gamma(x)}\,dx,
\ee
and $\gamma(x)$ is given by (\ref{gamma}). As is shown in
\cite{Zamolo}, the quantity $S^{(cl)}$ is essentially
the classical Liouville action computed on the canonical
Liouville field for a sphere with 3 conical singularities.

The conformal block $\cal F$ has a similar asymptotic:
\be
{\cal F}(\Delta_1,\ldots,\Delta_4,\Delta;x)=
e^{{1\over b^2}f(\eta_i,p;x)}.
\ee
Here $f(\eta_i,p;x)$ is the classical conformal block
and has the following power series expansion in $x$:
\be\label{class-f-expansion}
f(\eta_i,p;x)=(\delta-\delta_1-\delta_2)\log{x}+
{(\delta+\delta_1-\delta_2)(\delta+\delta_3-\delta_4)\over 2\delta} x
+O(x^2).
\ee
Here $\delta=p^2+1/4$ and $\delta_i=\eta_i(1-\eta_i)$. Unfortunately,
no closed form representation of the conformal block is known,
even for the classical block $f(\eta_i,p;x)$.

The probability is therefore given in $b\to 0$ limit by
the following expression:
\be
{\cal P}(p,x) = e^{-{1\over b^2}{\cal S}_{\eta,\ldots,\eta}(p|x,\bar{x})}.
\ee
Here ${\cal S}_{\eta,\ldots,\eta}(p|x,\bar{x})$ is the quantity
\be
{\cal S}_{\eta_1,\ldots,\eta_4}(p|x,\bar{x})=
S^{(cl)}(\eta_1,\eta_2,1/2+ip)+S^{(cl)}(\eta_3,\eta_4,1/2-ip)-
f(\eta_i,p,x)-f(\eta_i,p,\bar{x})
\ee
specialized to the case of all $\eta_i$ equal to $\eta$.
An extremum of ${\cal P}(p,x)$ as a function of $p$ at fixed $x$
is determined by the equation:
\be
{\partial\over\partial p}{\cal S}_{\eta_1,\ldots,\eta_4}(p|x,\bar{x})=0
\ee
specialized to the case $\eta_i=\eta$.
Using (\ref{S-cl}) this gives an equation:
\be\label{equation}
-i\log{\gamma^2(1-2ip)\over\gamma^2(1+2ip)}-
i\log{S_{\eta_1,\eta_2}(p)}-i\log{S_{\eta_3,\eta_4}(p)}=
-{\partial\over\partial p}\left( f(\eta_i,p;x)+f(\eta_i,p;\bar{x})\right),
\ee
where
\be
S_{\eta_1,\eta_2}(p)={\gamma(\eta_1+\eta_2-1/2+ip)
\gamma(1/2+\eta_1-\eta_2+ip)\over 
\gamma(\eta_1+\eta_2-1/2-ip)
\gamma(1/2+\eta_1-\eta_2-ip)}.
\ee
We will study this equation for the case of maximally massive
particles in the following section.

\section{Maximally massive particles: $\eta=1/4$}
\label{sec:massive}

As we shall see, a complete analysis is possible in this case in that
the dependence $\nu(x)$ can be obtained explicitly by finding
a conformal map for the Schottky uniformization, see previous
section. The obtained function $\nu(x)$ will agree with the relation obtained
by the second method, at least for small $x$, when it
is enough to keep only the logarithmic term on the right hand side
of (\ref{equation}). Our analysis will thus verify that the black hole 
creation probability is picked at the same Riemann surface 
as the one that arises in the analytic continuation of the classical 
collision process. This will solve problems (i) and (ii). The
problem (iii) of finding the probability as a function of the
modulus is dealt with in the next section.

\subsection{Extremum of the probability}

Let us first use the method based on the equation (\ref{equation}).
For $\eta=1/4$ it becomes:
\be\label{eq-massive}
2\pi-4i\log{\Gamma(1-2i p)\Gamma(1+ip)\Gamma(1/2+ip)\over 
\Gamma(1+2i p)\Gamma(1-ip)\Gamma(1/2-ip)}=
-{\partial\over\partial p}\left( f(1/4,p;x)+f(1/4,p;\bar{x})\right).
\ee
It should be understood as an equation on $p$ as a function of
$x$. Let us consider the case of small $x$. We shall assume that
this corresponds to small $p$, and later check that this assumption
is correct. For small $x$, the series representation (\ref{class-f-expansion}) 
of the classical conformal block gives $-2p\log{|x|^2}$ for the 
right hand side of (\ref{eq-massive}).
The small $p$ asymptotic of the left hand
side is obtained using the formula:
\be
{\rm Arg}\Gamma(x+iy)=y\Psi(x)+O(y^2).
\ee
Here $\Psi(x)=\Gamma'(x)/\Gamma(x)$.
The equation (\ref{eq-massive}) then becomes
$2\pi-16 p\log{2}=-2p\log{|x|^2}$, whose solution can be
written as:
\be\label{nu-x}
\nu(x)=2\pi p={\pi^2\over \log(1/|x|) + 4\log{2}}.
\ee
Thus, small $|x|$ indeed correspond to small $p$, as assumed.
We will verify this relation in the next subsection using a different
method.

\subsection{Uniformization map, relation to colliding particles}

In this subsection we find a map giving the
Schottky uniformization. This will allow us to determine
$\nu(x)$ explicitly. As we shall see, the map in question
is given by elliptic functions.

As we have described in the previous section, the Schottky uniformization
$w$ maps the Riemann surface $\tilde{X}=\C/\{0,x,1,\infty\}$
into a fundamental domain of some group $\Sigma$ on $\C$. Let
us take $\Sigma$ to be the group obtained in section \ref{sec:cont},
the one that corresponds to the zero impact parameter case, with 
$\eta=1/4$. A half $D$ of the fundamental 
domain for this case is shown in Fig.~\ref{fig:fund-region}(a).
It is enough to find a map from $D$ to the upper half-plane.
Such a map exists by Riemann's mapping theorem, and it is clear
that it coincides with the inverse of the map $w$.

We notice that one can first map
the unit disc into the upper half-plane by a
fractional linear transformation. One can then use the
logarithm function to map $D$ into a rectangle
of sides $\mu/\pi, i$, where 
$\mu$ is defined in (\ref{mu-two}). This rectangle
can be mapped into the upper half-plane by means
of elliptic function ${\rm sn}$.

\begin{figure}
\centerline{\hbox{\epsfig{figure=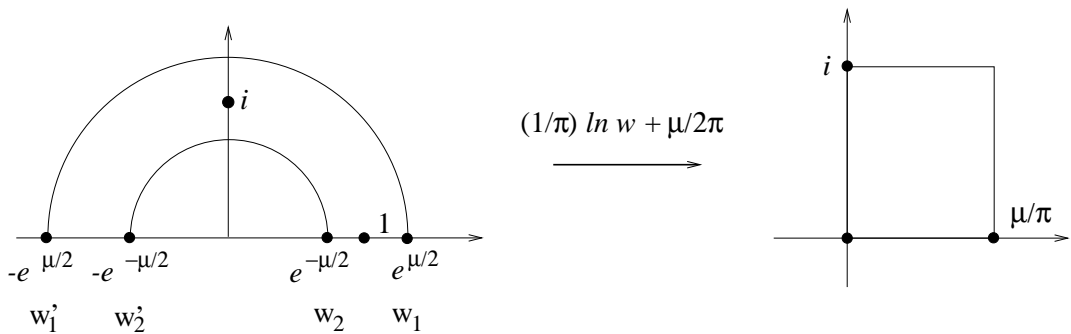,height=1.7in}}}
\caption{The unit disc can be mapped into the upper half-plane,
which is then mapped into an infinite strip by the logarithm
function. The images of $D$ are shown.}
\label{fig:D-images}
\end{figure} 

As before, we denote the local coordinate on $\tilde{X}$ by
$z$. The surface is then the complex $z$-plane with
points $0,x,1,\infty$ deleted. We choose the map $z(w)$ so that 
$w_2$ goes on the $z$-plane to $z_4=0$, point $w_1\to z_3=x, 
w_1'\to z_2=1$ and $w_2'\to z_1=\infty$. 
The map $z(w)$ is then given by:
\be\label{map-z}
z(w)=x\,{\rm sn}^2\,\left({\pi K(x) w\over\mu};\sqrt{x}\right),
\ee
with
\be\label{relation}
{K(x)\over K(1-x)}=\mu/\pi.
\ee
Here $K(x)$ is the usual complete elliptic integral of the first kind
given by
\be
K(m) = \int_0^1 {dt\over\sqrt{(1-t^2)(1-mt^2)}} = {\pi\over 2}
F(1/2,1/2,1;m),
\ee
and $F$ is the usual hypergeometric function. 

The relation (\ref{relation}) is the one we need. Indeed, in
the case $\eta=1/4$ the Riemann surface modulus $\nu=\mu$,
and thus $\nu(x)$ is given by (\ref{relation}). Let us
compare it with the relation we obtained in the previous
subsection. In the limit of small $x$
$K(x)\sim\pi/2$ and we only need $K(1-x)$. The corresponding 
asymptotic can be obtained
by using an expansion for $F(a,b,a+b,z)$. Keeping only the
zeroth order terms we get:
\be
F(1/2,1/2,1;1-x)={1\over \pi}\left(\log(1/x)+4\log{2}\right)+
O(x\log{x}).
\ee
It is now easy to see that in the limit of small $x$ the
formula (\ref{relation}) gives the same dependence
$\nu(x)$ as we have obtained earlier in (\ref{nu-x}).
This verifies that the probability is picked at the
same Riemann surface as the one obtained by the analytic
continuation.

The relation (\ref{relation}), although almost
obvious in the retrospect, seems to have not been noticed
before. What is known, see \cite{Zamolo-Conf-Block}, is that
\be\label{q}
q = \exp{-\pi K(1-x)/K(x)}
\ee
is a more natural parameter than $x$ for purposes of, e.g.,
developing a series representation for the conformal block.
However, the fact that $iK(x)/K(1-x)$ is the modulus of the
Riemann surface corresponding to $\eta=1/4$ seems new.

Let us note that although our derivation of the relation 
(\ref{relation}) used in an essential way the fact that
we are dealing with massive $\eta=1/4$ particles, for a large BH size 
$\nu$ one can expect (\ref{relation}) to hold for any $\eta$. 
Indeed, let us take the massless $\eta=1/2$ case.
A half of the fundamental region for this case is shown in 
Fig.~\ref{fig:fund-region}(b). It is clear that for
large $\nu$ the fundamental region becomes essentially the same as that
for $\eta=1/4$ large $\nu$, namely the whole unit circle.
This means that for large $\nu$ the map $w: D\to \U$ must
be insensitive to $\eta$. For large BH one should thus expect 
(\ref{relation}) to be valid independently of value of $\eta$. We
will use this in the next section when analyzing the
large BH creation probability.

\section{Probability for a large BH}
\label{sec:large}

To estimate the probability $\cal P$ for a large size
of the BH created we can utilize the result of
\cite{Zamolo-Al} that states that the conformal block
${\cal F}(\Delta_i,\Delta;x)$ is, for large $\Delta$,
given by:
\be
{\cal F}(\Delta_i,\Delta;x)\simeq q^\Delta,
\ee
where $q$ is given by (\ref{q}).
Using this, and the fact that the probability is given by the
4-point function, which is the absolute value square of
${\cal F}$, we see that:
\be\label{prob**}
{\cal P}\simeq e^{-\nu/2 b^2},
\ee
where we have used that $\Delta=(1+p^2)/b^2$ and $\nu=2\pi p$.

The result (\ref{prob**}) is what we need. In the remainder
of this section we give a derivation of this result using 
the classical Liouville
action. This will allow us to give an interpretation
to (\ref{prob**}) in terms of the Euclidean
gravity action on the instanton that describes the colliding
particles, see the beginning of the next section. 
The derivation will also help us understand the case of
a non-zero impact parameter.
 
Thus, the probability ${\cal P}(\nu,x)$ is given
by the 4-point function projected onto a particular intermediate
state. As we have discussed, in the semi-classical limit that
we are interested in the 4-point function is dominated 
by the exponential of the classical Liouville action, 
see (\ref{domin}). This thus gives the probability of the
most likely BH:
\be\label{prob-Liouv}
{\cal P}(\nu,x(\nu))=e^{-{1\over b^2}S_{Liouv}[\tilde{\varphi}]}.
\ee
Here $S_{Liouv}[\tilde{\varphi}]$ is the Liouville action
evaluated on the canonical field $\tilde{\varphi}$ that corresponds
to the Riemann surface of modulus $\nu$. The problem thus reduces
to that of evaluating the Liouville action. We will find 
$S_{Liouv}[\tilde{\varphi}]$ in the limit of a large BH
created. As we shall see, in this limit the result is 
independent of the mass of colliding particles.

Let us first analyze the massive case $\eta=1/4$.
We evaluate the Liouville action on the Schottky uniformization domain. 
Our analysis is essentially the same as in 
\cite{Zamolo-Conf-Block,Zamolo-Al}. First we need to to find the
map $J^{-1}$ relating the Schottky and Fuchsian uniformizations,
see (\ref{diagram}). This map can be obtained as a ratio of two
linearly independent solutions of the Fuchs equation on the
Schottky domain. The quadratic differential that appears in this
equation is given by (\ref{accessory}) plus the Schwarzian derivative
of the map $z(w)$ (\ref{map-z}). We thus get a Schroedinger 
equation on the $w$-plane, with a doubly-periodic potential, 
see \cite{Zamolo-Al}. This means that two linearly independent
solutions of the Fuchs equation on the Schottky domain are
of the form $e^{\pm \nu w/2K}f_{1,2}(w)$, where $f_{1,2}$ are
certain doubly-periodic functions. The map $J^{-1}$ is therefore
given by:
\be
J^{-1}(w)=e^{\nu w/K} f_1(w)/f_2(w).
\ee
Here we assumed that a half of
the fundamental domain for $\Sigma$ is a rectangle of sides
$K=K(x), iK'=iK(1-x)$, with $\nu/\pi=K/K'$. The constant
in the exponential is then selected so that the monodromy
one gets by going from $w=0$ to $w=K$ and back (that is the
monodromy around the operators at $z_4=0$ and $z_3=x$)
is equal to $\cosh{\nu}$. The derivative of the Liouville field 
(\ref{Liouv-Schottky}) is given, for large $\nu$, by: 
$\tilde{\varphi}_w\simeq \nu/K$. The Liouville action is thus:
\be
S_{Liouv}[\tilde{\varphi}]\simeq {1\over 4\pi}\int d^2x\, |\nu/K|^2.
\ee
The integral here is taken over the fundamental domain on the
$w$-plane. The fundamental domain is a parallelogram formed
by vectors $2K, iK'$. The modular parameter of this parallelogram is
$\tau=iK'/2K$. Its area is: $|2K|^2 {\rm Im}(\tau)$. We then obtain:
\be
S_{Liouv}[\tilde{\varphi}]\simeq \pi |\nu/\pi|^2 {\rm Im}(\tau).
\ee
This is the same result as the one obtained in \cite{Zamolo-Al}.
Let us now use it to get the probability. 
For purely imaginary $\tau$, which is the case of a non-rotating
BH, ${\rm Im}(\tau)=\pi/2\nu$, and we get (\ref{prob**}).
It is not hard to see that this result also holds in the rotating
case. Indeed, for $\eta=1/4$ $\nu=\mu$, and rotation is incorporated
by replacing $\mu\to\mu+2i\chi$, where $2\chi$ is the impact parameter.
It is easy to see that the probability is independent of $\chi$ and is
given by (\ref{prob**}). 

Although we were discussing the case $\eta=1/4$ it is
clear that for large $\nu$ nothing depends on $\eta$. Indeed,
the only input in the calculation of the preceding paragraph is
the map (\ref{map-z}). For $\eta\not=1/4$ the map $z(w)$ giving
the Schottky uniformization is not known. However, one can use 
$z(w)$ given by (\ref{map-z}) just as a change of variables for the
Fuchsian equation. This maps the Riemann surface 
$\tilde{X}=\C/\{0,x,1,\infty\}$
into a parallelogram on the $w$-plane. In the $\eta>>1$ limit 
the Liouville action can be evaluated as above, giving
(\ref{prob**}) for the probability. It is also clear that
the same expression (\ref{prob**}) with $\nu\to{\rm Re}(\nu)$ 
is valid in the rotating case. This means that the
probability $\cal P$ is independent of the impact parameter,
for large values of $\nu$ and in the range of the impact
parameter far enough from the value at which an extremal
BH is created. It would be of interest to find the
probability in the near-extremal region. We shall not
attempt this in the present paper.

Let now restore the dependence on physical constants. Using
(\ref{b}) we get:
\be\label{probability}
{\cal P}(\nu)\simeq e^{-\nu l/8 l_p}.
\ee
We note that the quantity in the exponential 
equals to the quarter of the BH entropy $S_{\rm BH}=2\pi r_+ /4 l_p$. 

\section{Discussion}
\label{sec:discussion}

The result (\ref{probability}) can be interpreted in a different way.
Let us recall, see \cite{Riemann}, that the value of the
appropriately regularized Einstein-Hilbert action on a
3-dimensional hyperbolic manifold that has $\tilde{X}$
as the conformal boundary is equal to $S_{Liouv}$ evaluated
on the canonical Liouville field corresponding to $\tilde{X}$.
The result (\ref{prob**}) was obtained by evaluating
$S_{Liouv}$ on the Liouville field corresponding to 
the geometry of $\tilde{X}$, with $\tilde{X}$ being the 
conformal boundary of $M$, analytic continuation of
the colliding particles spacetime. Thus (\ref{probability})
can also be interpreted as the exponential of the 
classical gravity action evaluated on the instanton
that describes the colliding particles.

We have only analyzed the simplest case of large BH size.
It is not hard to extend this analysis to small $\eta$,
using the relation $\nu(x)$ that can be obtained
for $\nu<<1$ from (\ref{equation}). We shall not attempt
this in the present paper.

We have only considered the case of small AdS curvatures,
which, in view of (\ref{b}), corresponds to the
semi-classical limit of the LFT. In this case one can
use the fact that the 4-point function is dominated by
the classical Liouville action. However, the expression
(\ref{prob*}) is clearly valid for any $b$, thus one
can analyze the BH production in the regime of large curvatures. 
This is when the effects of quantum become crucial and quantum gravity
is essential. The theory leads to definite predictions. 
We leave an analysis of the large curvature regime to the future.

Let us note just one complication that arises in the case of
small $l$. Namely, it only makes sense to talk about a spacetime
interpretation in the semi-classical limit, when there is
a Riemann surface dominating the 4-point function, and its
parameters can be identified with those of the colliding
particles. In the full-fledged quantum theory no such
identification is possible. This means that in the quantum
regime the best one has for the measure of the particle's
momentum is the cross-ratio $x$, and the only available
measure of the BH size is the conformal dimension of
a primary state $|P\rangle$. For a fixed $x$ the probability is no
longer expected to be peaked at a particular $P$. Note, however,
that it does not mean that the energy is not conserved,
for there is no longer a direct relation between $x$ and
the momentum of particles.

Let us now discuss the physical implications of (\ref{probability}).
First of all, we note that (\ref{probability}) can also
be interpreted as the probability of a BH of size
$2\nu$ to evaporate into a configuration of two
point particles of equal mass.\footnote{%
I am grateful to S.\ Solodukhin for suggesting to me this
interpretation.} It is then not at all surprising that
the answer we obtained is exponentially small. The exponential
suppression is that of Boltzmann type, as could be expected. 
Indeed, the factor in the exponential in (\ref{probability})
is equal to $-\beta M/2$, where $\beta$ and $M$ are the BH
inverse temperature and mass correspondingly. Thus, the
exponential factor is the one relevant for a thermal
emission of a particle of mass $M/2$, which is a rather
natural answer. 

We should now recall that the geometry of the BH created,
see Fig.~\ref{fig:t0-plane}, is just one of the possible
outcomes of the collision process. As we discussed in
section \ref{sec:cont}, the topology inside the
horizon may be more complicated, see Fig.~\ref{fig:t0-other}.
From the point of view of the outside observer all these
geometries are indistinguishable. They can thus be thought of as
BH's microstates. Let us adopt this interpretation. 
Then the quantity relevant for the outside observer is
is the total probability of forming a horizon,
given by the sum of probabilities for all possible topologies inside.
Let us analyze the structure of this sum.
One obtains the probability for other topologies as the
4-point function on a higher genus surface that is the 
double of the spatial slice. Thus, the geometry shown in 
Fig.~\ref{fig:t0-other}(b) 
results in a genus one surface, and 
geometry shown in  Fig.~\ref{fig:t0-other}(a) gives
a genus two surface. Therefore, the sum over possible
topologies inside the horizon is given by 
\be\label{sum-top}
\lower0.25in\hbox{\epsfig{figure=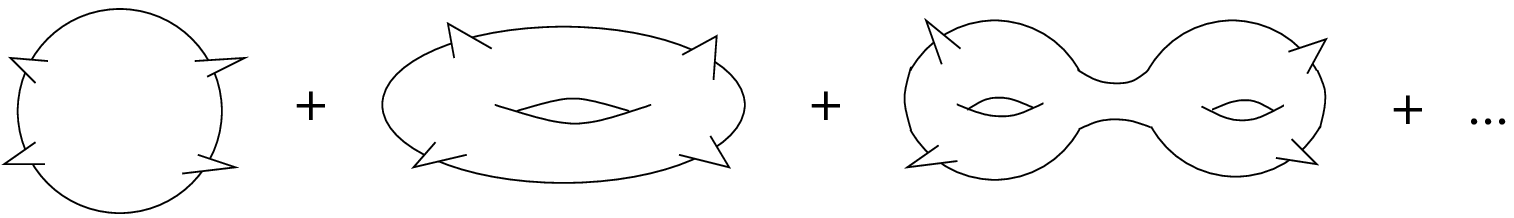,height=0.6in}}
\ee
Interestingly, one obtains the ``full'' 
string 4-point function given by a sum over the world-sheet genus.

\begin{figure}
\label{fig:t0-part}
\centerline{\hbox{\epsfig{figure=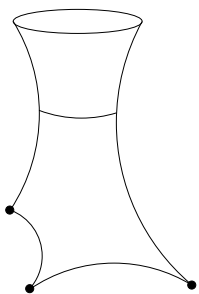,height=1.5in}}}
\caption{To get the total probability to form a horizon
one should in addition take into account processes in which
point particles split. For example, the two point
particles can split into three, as is shown here.}
\end{figure}

The above sum is not yet the total probability to create a horizon. 
The point is that, in addition to handles or other asymptotic
regions, the geometry inside the horizon may contain other
point particles. For example, in the geometry of 
Fig.~\ref{fig:t0-plane} one may have three point particles
inside the horizon instead of two, see Fig.~\ref{fig:t0-part}. 
This has an interpretation
of two particles colliding and producing a horizon, but then
splitting into 3 particles. Similarly, in the geometry of 
Fig.~\ref{fig:t0-other}(a)
the internal handle may degenerate into two point particles.
This would correspond to a process in which two point
particles produce a horizon, then splitting
into 4 point particles. To get the total probability
to form a horizon one has to add to (\ref{sum-top})
the probability of all these processes. One gets the
probability by forming a double Riemann surface. Thus,
the geometry in Fig.~\ref{fig:t0-part} gives a sphere with 6
conical singularities. To get the total probability one has to sum 
over the labels at 2 conical singularities:
\be\label{sum-part}
\sum_{\eta'} \lower0.25in\hbox{\epsfig{figure=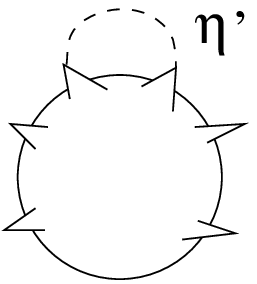,height=0.8in}}
\ee
One can similarly obtain probabilities for processes in which
more than one additional particle appears. The probability
is given by a $2n$-point function on a Riemann surface
of some genus, and one has to sum over all
different pairings of all vertex operators but those
corresponding to the 4 initial point particles. 

We thus see that the resulting
total probability of forming a horizon still has the structure of 
a sum over genus. However, one should now consider all $2n$-point
functions at a given genus, and then sum over all
different pairings of vertex operators, as in (\ref{sum-part}).
Note that this total probability can also be thought as
obtained from (\ref{sum-top}) in which one 
allows handles to degenerate into pairs of point particles,
and sums over the labels at these pairs. This gives a total
probability to form a horizon of any size. To get the probability of
creating a particular size horizon one should project
onto some intermediate state at one of the handles. 

The question now is whether the resulting sum 
still gives a small answer. A detailed answer to this
question would require understanding
of the second quantized LFT that governs the sum over
intermediate particle states in, e.g., (\ref{sum-part}), 
task beyond the scope of this paper. However, the simple
argument we gave in the Introduction, analog of the
argument of \cite{Voloshin}, tells us that we should
not expect an exponentially small production probability.
Let us discuss how an answer of order unity could result
from the sum over internal topologies in our theory.

First of all, it can be argued that
the sum in (\ref{sum-top}), without taking the possibility
of particle splitting into account, still gives an exponentially 
small answer.\footnote{%
I am grateful to M.\ Voloshin for explaining this to me.}
The point is that the other terms in the sum (\ref{sum-top})
can be expected to be even more suppressed. Indeed,
one expects that each handle introduces roughly the
factor of (\ref{prob**}). This is intuitively clear,
and is also supported by one's experience with 
exactly solvable models, in which sums like (\ref{sum-top})
can be exactly computed. For example, the multi-instanton
corrections in the sine-Gordon model result in the
probability given by \cite{Sine-Gordon}:
\be
{\cal P}_{multi} \sim \ln\left( 1- {\cal P} \right).
\ee
Here $\cal P$ is the exponentially small one-instanton
contribution. The right hand side, when expanded in powers
of ${\cal P}$, starts with ${\cal P}$ and is exponentially small.
All other terms are powers of ${\cal P}$ and are even more
suppressed. One can expect a similar behavior from the sum
(\ref{sum-top}). 

Thus, the sum (\ref{sum-top}) over internal topologies 
is unlikely to give any result different from (\ref{probability}).
However, the sum in which the internal geometry inside the
BH is allowed to have conical singularities may produce a
different answer. To see why this is so recall that, as
was argued in the companion paper \cite{String}, these
are the point particle states that are responsible for
the BH entropy. If this is so, then the sum over internal
particle states does produce a large factor of order $e^{S_{\rm BH}}$
necessary to make the total probability close to unity.

Another, independent way to argue in favor of this is as
follows. If the theory is unitary we must have that the
sum of probabilities of all possible outcomes, with and
without a horizon, must be one. The sum of probabilities
of all outcomes with a horizon is discussed above. The
outcomes in which no horizon is created are few: there
is the process in which two point particles just stick together,
or the process in which they scatter with no BH formed.
Let us first discuss the scattering process.
We first note that the 4-point function (\ref{4-point}) can be 
decomposed into intermediate states in a way different from
(\ref{prob}). Indeed, let us use the decomposition in a different
channel:
\be\label{interm-diff}
Z_{\rm LFT}\left[\,\,
\lower0.3in\hbox{\epsfig{figure=4-point.eps,height=0.6in}}
\,\,\right] = \sum_P \,\,
\lower0.34in\hbox{\epsfig{figure=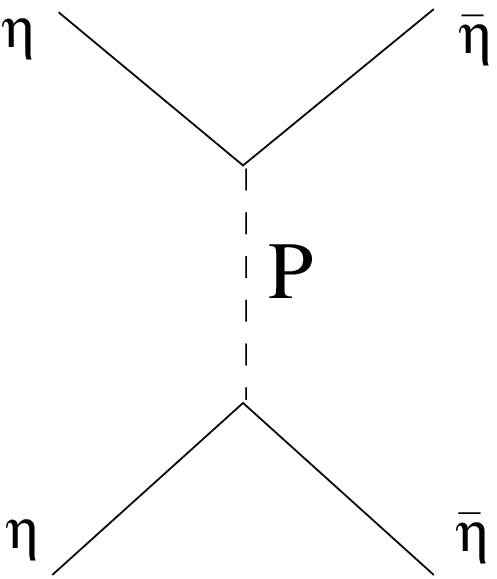,height=0.7in}}
\ee
Each term here can be interpreted as an amplitude for
particle scattering due to an exchange of a ``virtual'' BH
of size $P$. This is the same 4-point function as in
(\ref{4-point}), except that it is decomposed using a different
channel. It is thus also exponentially small, for values
of $x$ that give a large BH in the channel in (\ref{prob}).

There is also another scattering process. Namely,
there is a process given, similarly to (\ref{interm-diff}), by
an exchange of a virtual state, except that now this state
is that of a point particle:
\be\label{interm-part-diff}
\sum_{\eta'}\,\,
\lower0.34in\hbox{\epsfig{figure=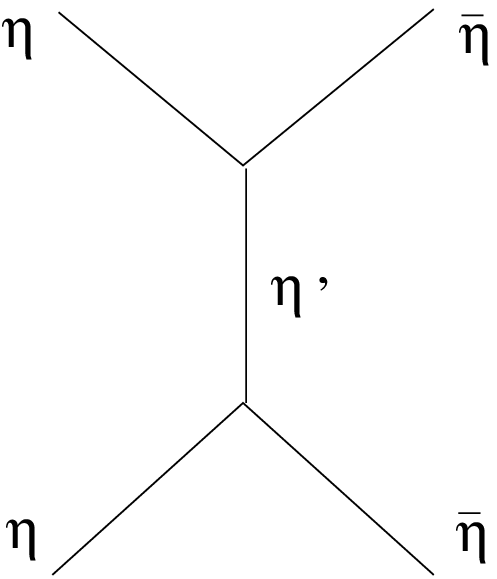,height=0.7in}}=
\sum_{\eta'} Z_{\rm LFT}\left[\,\,
\lower0.34in\hbox{\epsfig{figure=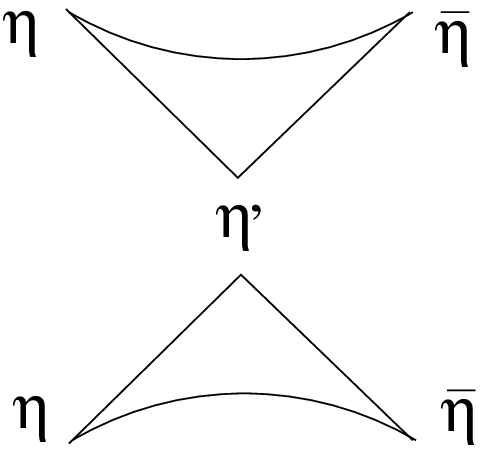,height=0.7in}}
\,\,\right].
\ee

The probability of two particles to stick together is closely
related to the one given above. In fact, it is given by the
same 4-point function decomposed into intermediate point
particle states, except that a different channel is used:
\be\label{interm-part}
\sum_{\eta'} Z_{\rm LFT}\left[\,\,
\lower0.25in\hbox{\epsfig{figure=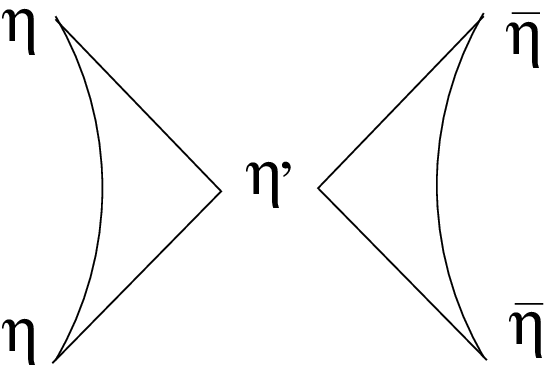,height=0.6in}}
\,\,\right].
\ee
The sum here should be equal to the sum over intermediate states
in (\ref{interm-part-diff}), in view of the crossing symmetry.
It is now not hard to argue that (\ref{interm-part}) must be
small. Indeed, the 4-point function projected onto an
intermediate state gives the probability to find
this state as an outcome of the collision. This probability is
expected to be picked at the process that is
realized classically. We have seen that this is indeed the
case for the BH intermediate state in (\ref{prob}). The
probability is picked at the BH that would be created classically.
Since for the range of $x$ that corresponds to a large BH
the process that is realized classically is creation of 
a BH, the probability of getting a point particles
as an outcome should be even smaller than (\ref{prob**}).
The sum in (\ref{interm-part}) should be thus expected to be
very small, and so should be (\ref{interm-part-diff}).
This implies that the probability of finding no horizon in the
final state is small. Thus, the total probability
of creating a horizon must be of order unity.

Let us summarize. We see how amplitudes (probabilities)
for every possible outcome of the collision process can
be computed. Our arguments indicate that the total
BH production rate is of order unity. To further test
this one would have to learn how to compute the sum over
internal particle states. It is
clear that this would require a much more detailed
description of the second quantized LFT than we have given 
so far. These, and other questions that remained
open we leave to the future work on the subject. 

\noindent
{\large \bf Acknowledgments}

I would like to thank R.\ Emparan, J.\ Hartle, G.\ Horowitz, M.\ 
Srednicki for discussions,
and S.\ Solodukhin for correspondence. Special thanks are to 
M.\ Voloshin discussions with whom were very useful.
The author was supported by the NSF grant PHY00-70895.

\eject
\appendix
\section{Appendix}
\label{app}

Unless specified otherwise we use units 
$8\pi G=\hbar=c=1$ throughout the paper.

Here we review some useful facts about 
the Lorentzian AdS${}_3$. This material is widely known, see,
e.g., \cite{Rot}. We need these facts in section \ref{sec:cont}.

The Lorentzian AdS${}_3$ can be defined as a
quadric $-U^2-V^2+X^2+Y^2=-l^2$ in $\R^{2,2}$. The metric is
given by: $ds^2=-dU^2-dV^2+dX^2+dY^2$. Thus, its group
of isometries is ${\rm O}(2,2)$. It is often more convenient to
introduce another set of coordinates $t,\rho,\theta$ defined by:
\begin{eqnarray*}\label{t-rho-theta}
U={1+\rho^2\over 1-\rho^2}\cos{t}, &\qquad&
V={1+\rho^2\over 1-\rho^2}\sin{t} \\
X={2\rho\over 1-\rho^2}\cos{\theta}, &\qquad&
Y={2\rho\over 1-\rho^2}\sin{\theta}.
\end{eqnarray*}
The metric then takes the following simple form:
\begin{equation}
ds^2 = - \left( {1+\rho^2\over 1-\rho^2} \right)^2 dt^2 +
\left( {2\over 1-\rho^2} \right)^2 (d\rho^2+\rho^2d\theta^2).
\end{equation}
Note that the constant $t$ planes in this model are all isometric to
the hyperbolic plane (in the Poincare unit disc model). The
conformal infinity $\cal I$ is the (timelike) unit cylinder $t,\theta$.

Another very convenient model, the one best suited for calculations,
is that of the $\SL(2,\R)$ group manifold, or, more precisely,
its universal cover. Note that the equation of the quadric
defining AdS${}_3$ can be rewritten as a requirement that
the following $2\times 2$ matrix has the unit determinant:
\begin{equation}
{\bf x} = \left( \begin{array}{cc}
  U+X & Y+V \\
  Y-V & U-X \end{array} \right)
\end{equation}
This makes it clear that AdS${}_3$ can be realized as the universal
cover of the $\SL(2,\R)$ group manifold. In this model the
metric is just the natural metric on the group manifold:
\begin{equation}
ds^2 = {1\over2}{\rm Tr}({\bf x}^{-1} d{\bf x} {\bf x}^{-1} d{\bf x}).
\end{equation}
This model also makes it clear that isometries can
be realized as the left and right action of $\SL(2,\R)$, that
is 
\be
{\bf x} \to A^L {\bf x} {A^R}^{-1}.
\ee

Let us also give a parameterization of the $\SL(2,\R)$ group
manifold in terms of coordinates the $t,\rho,\theta$ introduced
in (\ref{t-rho-theta}). We have:
\begin{equation}\label{point}
{\bf x} = {1+\rho^2\over 1-\rho^2} {\bf \omega}(t) +
{2\rho\over 1-\rho^2} {\bf \gamma(\theta)},
\end{equation}
where
\begin{equation}
{\bf \omega}(\alpha) = \cos(\alpha) {\bf 1} + \sin(\alpha){\bf \gamma}_0,
\qquad
{\bf \gamma}(\alpha) = \sin(\alpha) {\bf \gamma}_1 +
\cos(\alpha){\bf \gamma}_2,
\end{equation}
and $\gamma_a$ are the $\gamma$-matrices in 2+1 dimensions:
\begin{eqnarray}\label{gamma-matrices}
\gamma_0 = \left(\begin{array}{cc}
0 & 1 \\ -1 & 0 \end{array}\right), \qquad
\gamma_1 = \left(\begin{array}{cc}
0 & 1 \\ 1 & 0 \end{array}\right), \qquad
\gamma_2 = \left(\begin{array}{cc}
1 & 0 \\ 0 & -1 \end{array}\right).
\end{eqnarray}
These matrices satisfy:
\begin{equation}
{\bf \gamma}_a {\bf \gamma}_b = \eta_{ab} {\bf 1} - \varepsilon_{ab}^c\gamma_c,
\end{equation}
where $a,b = 0,1,2$, $\eta_{ab}={\rm diag}(-1,1,1)$ is the
three-dimensional Minkowski metric which is used to raise and
lower indices, and $\varepsilon^{abc}$ is the Levi-Civita symbol
with $\varepsilon^{012}=1$.

We also need an expression for isometry generating VF's.
Let us denote $J_{XY}=X\partial_Y-Y\partial_X,
J_{YV}=Y\partial_V+V\partial_Y$, etc. The two sets of 
commuting VF's, and their restriction to
the boundary $\cal I$ is given by:
\begin{eqnarray}\nonumber
J_1 = -{1\over2}(J_{XU}+J_{YV})=\sin{u}\partial_u, &\qquad&
\tilde{J}_1 = -{1\over2}(J_{XU}-J_{YV})=\sin{v}\partial_v, \\ 
\label{vf-boundary}
J_2 = -{1\over2}(J_{XV}-J_{YU})=-\cos{u}\partial_u, &\qquad&
\tilde{J}_2 = -{1\over2}(J_{XV}+J_{YU})=-\cos{v}\partial_v, \\ \nonumber
J_3 = -{1\over2}(J_{XY}-J_{UV})=\partial_u, &\qquad&
\tilde{J}_2 = {1\over2}(J_{XY}+J_{UV})=\partial_v.
\end{eqnarray}
This table is from \cite{Rot}. The VF $J_i, \tilde{J}_i$
are generators of the two copies of the Lie algebra ${\mathfrak sl}(2)$.
Here we have also indicated what these VF
become on the conformal boundary cylinder $\cal I$: $u=t-\phi, v=t+\phi$
are the usual null coordinates on $\cal I$. Let us also
note that the generators $J_i$ can be expressed in terms of the
$\gamma$-matrices. We have:
\begin{eqnarray}\label{J-gamma}
J_1 = -{1\over 2}\gamma_1, \qquad
J_2 = -{1\over 2}\gamma_2, \qquad
J_3 = -{1\over 2}\gamma_0.
\end{eqnarray}

\newcommand{\hep}[1]{{\tt hep-th/{#1}}}
\newcommand{\gr}[1]{{\tt gr-qc/{#1}}}

\end{document}